\documentclass[final,5p,times,twocolumn]{elsarticle}

\usepackage{amssymb}
\usepackage{amsthm}
\usepackage{amsmath}
\usepackage{array}
\usepackage{graphicx}
\usepackage{theorem}
\usepackage{dcolumn}
\usepackage{bm}
\usepackage{setspace}
\usepackage{stfloats}
\usepackage{subcaption}
\usepackage{caption}
\usepackage{textgreek}
\usepackage{xcolor}

\biboptions{sort&compress}

\usepackage{xtab}
\xentrystretch{-0.1}

\usepackage{lineno}

\begin{document}

\begin{frontmatter}



\title{Advances in understanding vacuum break dynamics in liquid helium-cooled tubes for accelerator beamline applications}


\author[label1,label2]{Yinghe Qi}
\author[label1,label2]{Wei Guo\corref{c1}}
\address[label1]{National High Magnetic Field Laboratory, 1800 East Paul Dirac Drive. Tallahassee, Florida 32310, USA}
\address[label2]{Mechanical Engineering Department, FAMU-FSU College of Engineering, Florida State University, Tallahassee, Florida 32310, USA}
\cortext[c1]{Corresponding author: wguo@magnet.fsu.edu}

\begin{abstract}
Understanding air propagation and condensation following a catastrophic vacuum break in particle accelerator beamlines cooled by liquid helium is essential for ensuring operational safety. This review summarizes experimental and theoretical work conducted in our cryogenics lab to address this issue. Systematic measurements were performed to study nitrogen gas propagation in uniform copper tubes cooled by both normal liquid helium (He I) and superfluid helium (He II). These experiments revealed a nearly exponential deceleration of the gas front, with stronger deceleration observed in He II-cooled tubes. To interpret these results, a one-dimensional (1D) theoretical model was developed, incorporating gas dynamics, heat transfer, and condensation mechanisms. The model successfully reproduced key experimental observations in the uniform tube system. However, recent experiments involving a bulky copper cavity designed to mimic the geometry of a superconducting radio-frequency (SRF) cavity revealed strong anisotropic flow patterns of nitrogen gas within the cavity, highlighting limitations in extrapolating results from simplified tube geometries to real accelerator beamlines. To address these complexities, we outline plans for systematic studies using tubes with multiple bulky cavities and the development of a two-dimensional (2D) model to simulate gas dynamics in these more intricate configurations. These efforts aim to provide a comprehensive understanding of vacuum breaks in particle accelerators and improve predictive capabilities for their operational safety.
\end{abstract}


\end{frontmatter}


\section{Introduction}
A sudden loss of vacuum (SLV) in cryogenic systems, including storage vessels, Nuclear Magnetic Resonance (NMR) systems, and particle accelerators, presents a significant safety risk due to the potential for rapid and dangerous pressure buildup \cite{padamsee2015design, harrisonLossVacuumExperiments2002, petitpasModelingSuddenHydrogen2013, heidtModelingPressureIncrease2014, xieStudyHeatTransfer2010}. Specifically, in particle accelerators, which are often composed of interconnected superconducting radio-frequency (SRF) cavities housed within cryomodules, vacuum failure can be particularly catastrophic. The long liquid helium (LHe) cooled beamline tube, collectively formed by SRF cavities, can be susceptible to air infiltration during such a failure. This air can condense on the inner surfaces of the beamline tube, depositing heat to the LHe and causing violent boiling, which may result in hazardous pressure buildup within the cryomodule \cite{adyLeakPropagationDynamics2014,bajkoReportTaskForce2009,seidelFailureAnalysisBeam2002,wisemanLossCavityVacuum1994,wiseman1991cebaf}. Additionally, air propagation through the beamline can lead to contamination with dust and other particles, potentially deviating the high cleanliness standards required for superconducting cavities \cite{wisemanLossCavityVacuum1994}. Understanding the complex heat and mass transfer processes coupled with vacuum break in a long LHe-cooled vacuum tube is crucial for designing effective safety components and mitigating the associated risks.

The behavior of gas propagation in such cryogenic environments is markedly different from that in room-temperature conditions. At room temperature, the speed of the gas shock wave, known as the "escape speed", is well understood and can be approximated using the relationship $v_0 = \frac{2c_0}{\gamma - 1}
$ , where $\gamma$ is the ratio of the specific heats and $c_0$ is the speed of sound in the gas. The typical propagation speeds can reach up to 1655 m/s \cite{toroRiemannSolversNumerical2013,shapiroDynamicsThermodynamicsCompressible1953}. However, in cryogenic systems, particularly those using LHe, the situation is different due to the process known as cryopumping, where gas molecules condense or freeze onto the cold surfaces of the tube, dramatically slowing down the propagation of the gas front \cite{bosqueTransientHeatTransfer2014,welchCapturePumpingTechnology2001,dawsonCryopumping1965,rogersExperimentalInvestigationsSolid1966,tantosGasFlowAdsorbing2016,brownCondensation3002500Gases1970}.

Studies conducted at Fermi National Accelerator Laboratory and the European X-ray Free-Electron Laser (XFEL) have observed that the gas front propagation speed in LHe-cooled beamline tubes is significantly slower—on the order of 10 m/s—compared to room-temperature tubes \cite{boeckmannExperimentalTestsFault2008,dalesandroResultsSuddenLoss2014}. Early experiments by Dhuley and Van Sciver in our lab, using straight tubes immersed in normal liquid helium (He I), demonstrated an exponential slowing effect, which they attributed to gas condensation \cite{dhuleyPropagationNitrogenGas2016,dhuleyPropagationNitrogenGas2016a}. Additionally, their preliminary results indicated a stronger slowing effect in tubes cooled by superfluid helium (He II)~\cite{dhuleyGasPropagationLiquid2016}. However, detailed quantitative modeling was not initially provided.

Building on this foundation, we have conducted systematic studies in recent years to advance the understanding of vacuum breaks in LHe-cooled tubes. Improvements to the experimental system—such as the addition of a vacuum jacket, multi-layer insulation (MLI), and the replacement of the straight tube with a much longer helical design—enabled precise control of the nitrogen gas condensation point and facilitated further investigation into the deceleration of gas propagation \cite{garceauDesignTestingLiquid2019, garceauHeatMassTransfer2019, garceauEffectMassFlow2020}. A comprehensive one-dimensional (1D) theoretical model was developed, systematically addressing gas propagation and condensation dynamics, and was validated against experimental data. Numerical simulations based on this model provided valuable insights into the complex heat and mass transfer processes, particularly the impact of mass deposition on gas propagation, which in turn influences heat transfer to both the He I and He II baths \cite{baoHeatMassTransfer2020, garceauHeatMassTransfer2021}. Our recent work expands upon these findings by incorporating a bulky cavity into the original uniform tube \cite{garceau2022vacuum}. Preliminary experimental results have revealed the anisotropic nature of gas flow inside the cavity, highlighting the challenges and needs for further systematic investigation posed by non-uniform tube configurations, commonly encountered in real accelerator beamlines containing volumetric SRF cavities.

This paper begins with a brief overview of previous experimental studies on vacuum failure in LHe-cooled beamline tubes, presented in Section 2. Section 3 then delves into our laboratory’s systematic research, detailing key findings from both experimental and theoretical perspectives. Finally, Section 4 highlights prospective research directions for our lab, introducing recent experimental progress and outlining plans for future work aimed at addressing existing challenges and advancing the understanding of vacuum failure dynamics.

\section{\label{sec:prior study}Prior studies}

Sudden vacuum loss represents a significant hazard to accelerator beamlines, prompting extensive experimental investigations at accelerator laboratories worldwide to explore the associated fault conditions \cite{wiseman1991cebaf, wisemanLossCavityVacuum1994, boeckmannExperimentalTestsFault2008, adyLeakPropagationDynamics2014, dalesandro2012experiment, dalesandro2014results}. These studies have primarily focused on the effects of vacuum breaches on RF performance and heat flux to the liquid helium bath. At the Continuous Electron Beam Accelerator Facility (CEBAF), experiments simulated vacuum loss in a quarter cryomodule containing two SRF cavities immersed in 2 K and 4 K LHe. These tests verified the effectiveness of cryomodule pressure relief systems and vacuum interlocks. Heat fluxes to the helium bath were calculated to reach sustained levels of 20 kW/m², with peak values of 35.0 kW/m² in He II and 28.4 kW/m² in He I \cite{wisemanLossCavityVacuum1994}.
Similarly, at Deutsches Elektronen-Synchrotron (DESY), a cryomodule was tested on the Cryomodule Test Bench (CMTB) to evaluate vacuum break fault conditions. The results indicated that insulation vacuum venting had a minimal impact on RF and cryogenic performance, whereas cavity vacuum venting significantly degraded RF performance. While heat transfer to the helium bath was observed, its estimation was subject to considerable uncertainty (±50\%) due to temperature inhomogeneities. Notably, air propagation along cavity surfaces was unexpectedly slow, with a measured pressure propagation speed of 3 m/s, taking 4 seconds to traverse the 12-meter cryomodule \cite{boeckmannExperimentalTestsFault2008}. This remarkable deceleration in propagation speed garnered significant attention, leading to further investigations into the underlying phenomena.
At Fermilab, an experimental system was developed to simulate sudden vacuum loss in a scaled Project-X SRF cryomodule, focusing on longitudinal pressure effects \cite{dalesandroResultsSuddenLoss2014, dalesandroExperimentTransientEffects2012}. In these experiments, which were conducted using a 2-meter-long beam tube with a 25-mm diameter, cryopumping effects were clearly observed, resulting in pressure propagation velocities of approximately 10 m/s, consistent with the findings at DESY \cite{boeckmannExperimentalTestsFault2008}. Interestingly, the temperature profile was found to propagate faster than the pressure profile, providing valuable insights into the complex dynamics of vacuum loss events.

To deepen the understanding of the complex heat and mass transfer processes involved in a beamline vacuum break, Dhuley and Van Sciver from our cryogenics lab conducted pioneering experiments \cite{dhuleyPropagationNitrogenGas2016,dhuleyPropagationNitrogenGas2016a,dhuleySuddenVacuumLoss2014,dhuley2014cryodeposition,dhuley2015heat}. They vented room-temperature nitrogen gas from a buffer tank into a LHe-cooled straight vacuum tube, 1.5 m in length with an inner diameter of 32 mm.
Their experiments with He I revealed that the propagation of the gas front decelerated nearly exponentially, a behavior attributed to gas condensation on the tube walls. However, the physical mechanisms underlying this exponential deceleration were not fully elucidated. While a simple analytical model was proposed to describe the phenomenon, it lacked the ability to quantify propagation speed due to the absence of a detailed description of the mass deposition rate.
Dhuley further explored this phenomenon using He II and reported preliminary findings in his dissertation, which suggested that He II might induce a stronger slowing effect than He I \cite{dhuleyGasPropagationLiquid2016}. Despite these insights, a more systematic investigation remains necessary. The experimental setup using a straight tube proved inadequate for precise He II measurements because, after the helium bath is pumped to the superfluid phase, only a small portion of the tube remains immersed in He II. Furthermore, the section of the tube above the liquid level is cooled by helium vapor to temperatures low enough to cause nitrogen gas to condense at an indeterminate location. This lack of precision adds complexity to data analysis and interpretation.
To address these challenges, a better controlled tube system and a comprehensive theoretical model are required to quantitatively investigate gas propagation dynamics under both He I and He II conditions. These improvements provide the precision necessary to fully characterize the intricate processes at play.

\section{\label{sec:our recent progress}Comprehensive investigations by our cryogenics lab}

Building on prior work by accelerator laboratories and the studies conducted by Dhuley and Van Sciver, our research seeks to systematically investigate the heat and mass transfer processes and mechanisms involved in a sudden vacuum break within a liquid helium-cooled beamline tube. The primary objectives include implementing experimental setup modifications to precisely control the onset of condensation, conducting systematic measurements to analyze longitudinal effects, and developing comprehensive theoretical models to simulate the intricate heat and mass transfer dynamics.

\subsection{Experimental system modifications}\label{sec:exp_setup}

To overcome the limitations of earlier vacuum break studies conducted with a straight copper tube immersed in LHe, a new experimental setup was designed and implemented to more effectively capture gas propagation dynamics during vacuum loss events. The detailed system modifications are thoroughly described by Garceau et al. \cite{garceauDesignTestingLiquid2019}.

\begin{figure}[!tb]
	\centering
	\includegraphics[width=\columnwidth]{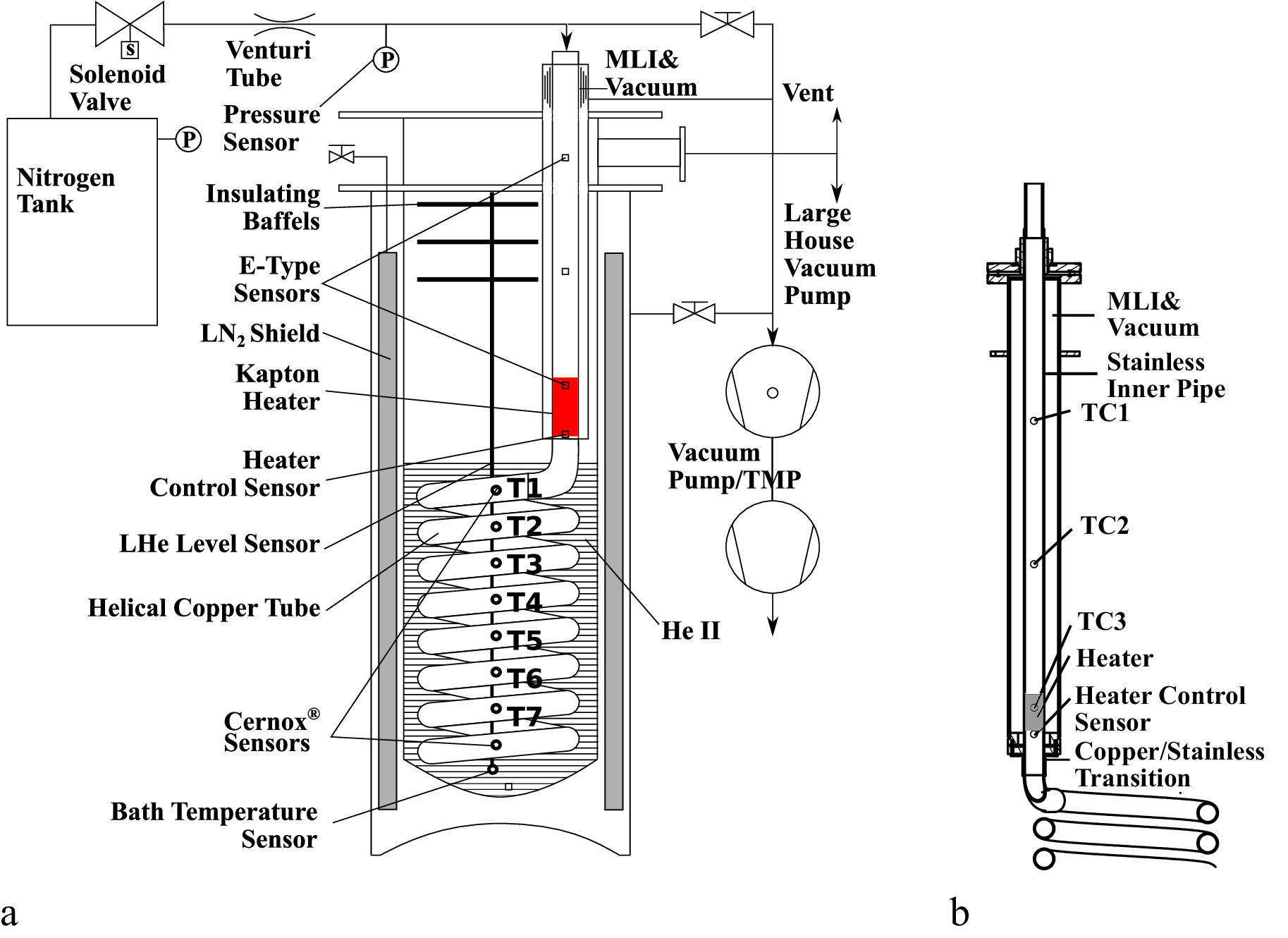}
	\caption{Schematics of (a) the updated helical tube system and (b) the enlargement of the upgraded vacuum jacket, copied from \cite{garceauDesignTestingLiquid2019, garceauHeatMassTransfer2021}.
	\label{fig:fig1}}
\end{figure}

The previous system, developed by Dhuley and Van Sciver, encountered challenges during He II experiments due to evaporative cooling, which left a significant portion of the tube unimmersed in liquid helium. This limitation hindered the accurate characterization of the slowing gas front \cite{dhuleyPropagationNitrogenGas2016,dhuleyPropagationNitrogenGas2016a,dhuleyGasPropagationLiquid2016}. To address this issue, the system was upgraded to incorporate a copper helical tube, coiled with a diameter of 22.9 cm and a pitch of 5.1 cm, which extended the tube length from 1.5 m to 5.75 m. Additionally, the buffer tank volume was increased from 86 L to 230 L to support extended measurements, as shown in Fig.~\ref{fig:fig1}a. A comparison of key specifications between the upgraded helical tube system and Dhuley and Van Sciver's original straight tube system is presented in Table~\ref{Ta:SysCompare}. The redesigned system facilitated more comprehensive gas propagation measurements over longer distances and durations, particularly under He II conditions \cite{garceauHeatMassTransfer2019}.

\begin{table}[h]
			\begin{center}
				\caption{Key specifications  of Dhuley and Van Sciver's straight tube system and the upgraded helical tube system, copied from \cite{garceauHeatMassTransfer2019}.}\label{Ta:SysCompare}
				\begin{tabular}{l c c}
					\hline
														& Straight  	& Helical  \\
					\hline
					Copper tube length (m) 				& 1.5			& 5.75\\
					Inner tube diameter (mm)				& 31.1			& 25.4	\\
					Tube wall thickness (mm)& 3				& 1.25 	\\
					Coil diameter (mm)					& --			& 229 	\\
					Coil pitch (mm)						& -- 			& 51\\
					N\textsubscript{2} reservoir (L)	& 86			& 230 	\\
					\hline
				\end{tabular}
			\end{center}
		\end{table}

To further address excessive gas slowing in the helical tube during He II experiments, a stainless steel vacuum shield with multilayer insulation (MLI) was installed around the 2.8 cm stainless steel extension pipe connecting the external plumbing to the copper helical tube, as shown in Fig.~\ref{fig:fig1}b. Additionally, a Kapton tape heater was added to maintain the temperature above nitrogen’s condensation point (77 K at 1 atm \cite{nist_nitrogen_2019}), with precise regulation provided by a Lake Shore 340 Temperature Controller. The controller was equipped with a sensor positioned 35 mm above the copper-to-stainless steel transition. To ensure stable experimental conditions, the temperature profile of the inner tube’s upper section was monitored using thermocouples placed at 65 mm, 265 mm, and 465 mm above the transition point, enabling accurate measurements of gas propagation dynamics.

To investigate vacuum break scenarios, high-purity dry nitrogen gas (99.999\%) from the 230-liter buffer tank was used as a substitute for air. This choice eliminated the variability associated with air’s mixed composition of nitrogen, oxygen, and trace components. Vacuum breaks were initiated by activating a solenoid valve with a rapid opening time of 25 milliseconds, allowing nitrogen to flow into the evacuated beam tube. To control the flow, a venturi pipe positioned immediately downstream of the solenoid valve was used, ensuring the gas exited at sonic velocity.

Upon exiting the venturi, the gas passed through a temperature-controlled, vacuum-insulated section of stainless steel tubing before entering the copper helical coil submerged in liquid helium. Surface temperatures along the copper coil were monitored at eight points using Lake Shore Cernox\textsuperscript{\textregistered} sensors embedded in 2850 FT Stycast\textsuperscript{\textregistered} epoxy, which provided insulation against the surrounding LHe environment \cite{dhuley2016epoxy,dhuleyPropagationNitrogenGas2016}. The first temperature sensor was mounted 6 cm downstream from the copper elbow marking the transition from the vacuum-jacketed section, with additional sensors spaced at 72 cm intervals along the helical tube. To improve sensor performance, the copper surface was polished at mounting points, and thermal resistance was minimized using indium foil and Apiezon\textsuperscript{\textregistered} N thermal grease. Stainless steel clamps secured the sensors in place.

Pressure measurements were conducted to complement the temperature readings. The inlet vacuum pressure of the helical tube was monitored using a cold cathode gauge, capable of detecting pressures from $10^{-3}$ to $10^{-7}$ Torr. Simultaneously, the nitrogen supply pressure was measured using a 1000 Torr MKS 626-Baratron\textsuperscript{\textregistered} capacitance manometer and a high-speed Kulite\textsuperscript{\textregistered} XCQ-092 pressure sensor. These sensors, combined with calculations based on the methodology established by Dhuley and Van Sciver \cite{dhuleyPropagationNitrogenGas2016}, provided valuable insights into the nitrogen mass flow rate. Furthermore, an additional Kulite\textsuperscript{\textregistered} sensor positioned downstream of the venturi captured supplementary gas flow properties, enabling a more comprehensive analysis of the system.

Data collection utilized three Keithley 2700 Data Acquisition/Multimeter systems for temperature readings and four DT9824 USB data acquisition modules from Data Translation, Inc., which recorded high-speed temperature and pressure data at 4800 Hz. These modules were synchronized through a voltage signal from the solenoid valve to establish a common time-zero reference. The experimental system was controlled and monitored using National Instruments LabVIEW\textsuperscript{\textregistered} software, which handled sensor integration and data acquisition.

This upgraded experimental system enabled a more precise investigation of gas propagation dynamics during vacuum loss, particularly in the context of He II experiments, overcoming the limitations of earlier straight-tube configurations. The enhanced measurement capabilities, including improved temperature and pressure monitoring and gas composition control, provide a robust platform for studying vacuum break phenomena in complex cryogenic environments such as particle accelerators.

\subsection{Systematic measurements}
\subsubsection{Experimental procedure}\label{sec:exp_procedure}
In a typical experiment, the beam tube, pipe insulation jacket, and cryostat vacuum jacket were evacuated over two or more days using a turbo molecular pump system. Vacuum pumping was considered sufficient when the pressure reached the $10^{-5}$ Torr range. Once the target vacuum level was achieved, the system was precooled with liquid nitrogen (LN$_2$) by filling the nitrogen shield and the liquid helium (LHe) bath.

The following day, after the precooling phase, the LHe bath was drained of LN$_2$ and allowed to warm to 90 K to ensure the complete removal of residual LN$_2$ from the system. Subsequently, the LHe bath was refilled over several hours. For He I experiments, the bath was filled to a level corresponding to a 46 cm reading on the liquid level sensor (LLS), approximately aligning with the top of the copper elbow transitioning to the stainless steel pipe. This location also marked the lower end of the vacuum jacket in the upgraded system. For He II experiments, the LHe bath was filled to its maximum level, around 79 cm on the LLS, after which the system was vacuum pumped to evaporatively cool the helium from He I to He II. Following vacuum pumping, the liquid level in He II experiments was comparable to that in He I experiments, near the 46 cm mark on the LLS, with the temperature dropping below 2 K.

Once the desired LHe level and temperature were achieved, the tube system was isolated by closing all valves. Data acquisition commenced and was allowed to run for at least one second before initiating the vacuum break by opening the solenoid valve. The valve remained open for at least eight seconds before being closed. After the experiment, the system was left to safely boil off the remaining LHe and warm up, thereby preventing pressure buildup in the beam tube as the nitrogen returned to its gaseous state. Experiments were conducted at four different buffer tank pressures: 50 kPa, 100 kPa, 150 kPa, and 200 kPa.

\subsubsection{Key findings}
\begin{figure}[!tb]
	\centering
	\includegraphics[width=\columnwidth]{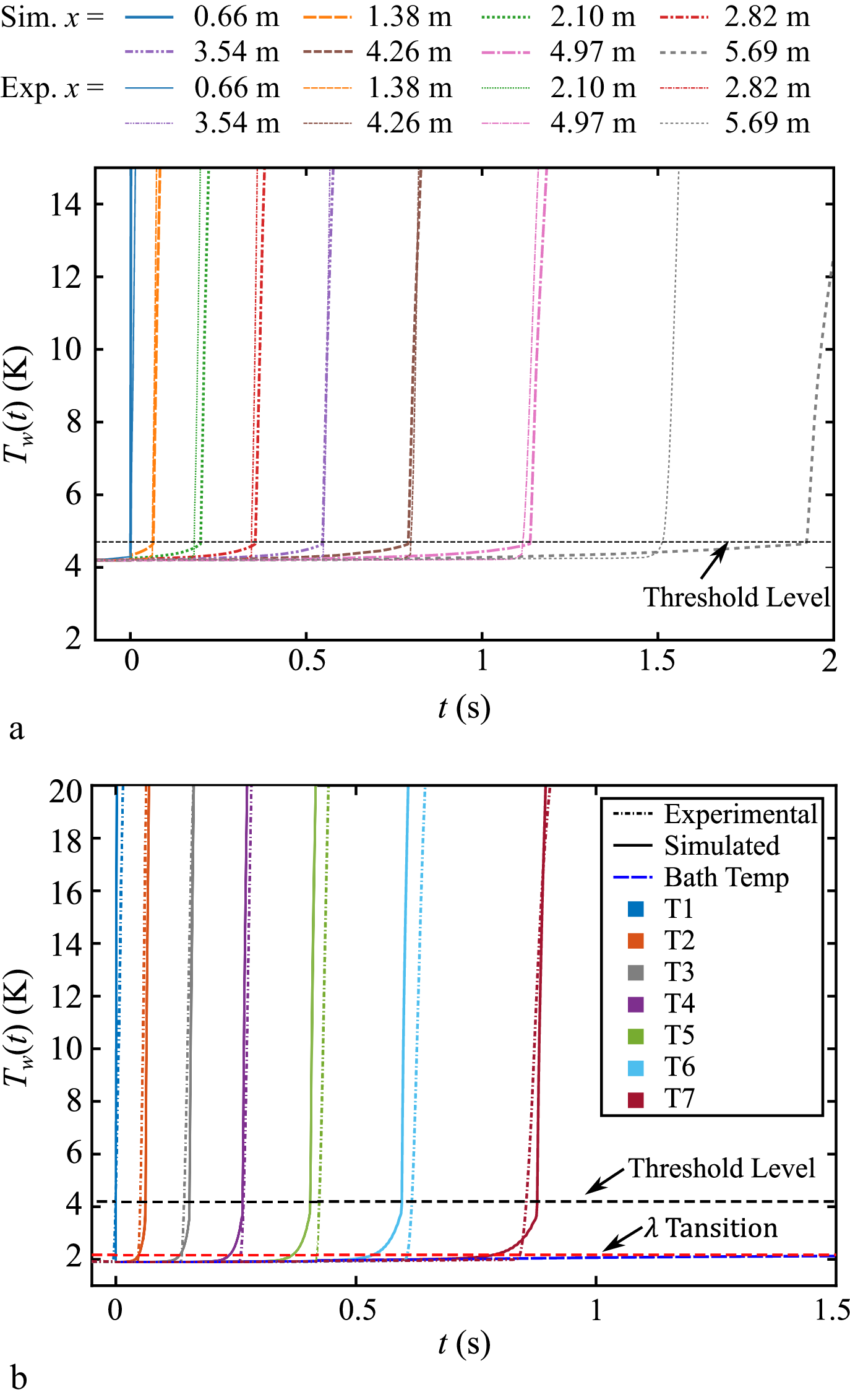}
	\caption{Comparison of the calculated and measured wall temperature curves $T_w(t)$ at locations where the thermometers are installed with (a) He I bath at a tank pressure of 100 kPa and (b) He II bath at a tank pressure of 150 kPa, copied from \cite{baoHeatMassTransfer2020,garceauHeatMassTransfer2021}.
	\label{fig:fig2}}
\end{figure}

Experiments conducted with the upgraded system, as detailed in previous studies \cite{garceauDesignTestingLiquid2019,garceauHeatMassTransfer2019,baoHeatMassTransfer2020,garceauHeatMassTransfer2021}, utilized temperature spikes on the tube wall, $T_w(x,t)$, observed after valve opening, to trace the gas front. To reduce random and harmonic noise, which varied across sensors and datasets, the sensor data were processed using an 80-point moving average filter for smoothing. Gas propagation was quantified by defining a rise time, $ t_r(x)$, as the point at which the temperature at a distance \( x \) from the coil entrance exceeded 4.7 K for He I and 4.2 K for He II (see Fig.~\ref{fig:fig2}). Although these thresholds were selected arbitrarily, tests confirmed that their precise values did not significantly affect model fitting results, provided they were chosen within the sharply rising portion of the temperature curves.

The dashed lines in Fig.~\ref{fig:fig2}a and b show representative wall temperature curves recorded by Cernox sensors during experiments with He I at 100 kPa and He II at 150 kPa, respectively. Initially, the wall temperature remains at the bath temperature (4.2 K for He I and 1.95 K for He II). As GN\(_2\) enters and condenses on the tube wall, the temperature rises sharply, peaking around 50–60 K (not shown in the figure), and then gradually declines due to SN\(_2\) frost formation on the wall. The increasing spacing between successive curves indicates a deceleration of the gas front. Simultaneously, heat deposited in the bath raises the bath temperature, $T_b$; however, in He II, \( T_b \) remains below the lambda point (2.17 K), maintaining He II phase throughout and allowing He II heat transfer characteristics to inform our analysis.

\begin{figure}[!tb]
	\centering
	\includegraphics[width=\columnwidth]{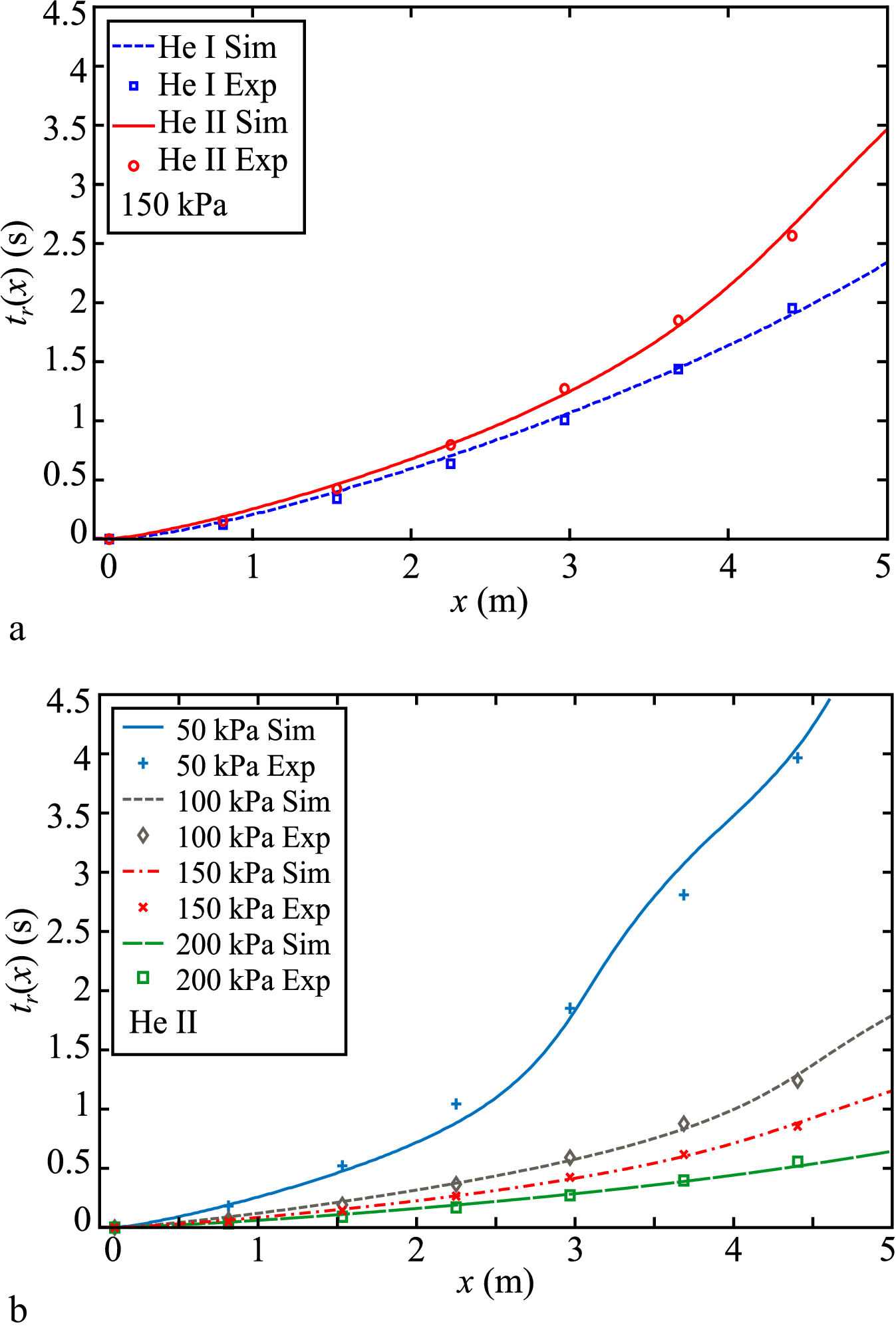}
	\caption{Measured and simulated rise times as a function of position for (a) He I and He II at a tank pressure of 150 kPa and (b) He II at different tank pressures, copied from \cite{garceauHeatMassTransfer2021}.
	\label{fig:fig3}}
\end{figure}

Fig.~\ref{fig:fig3} presents the rise time, $t_r(x)$, as a function of position, $x$. In Fig.~\ref{fig:fig3}a, the rise time curves for He I and He II are displayed at an equal tank pressure of 150 kPa, with He II demonstrating a markedly stronger deceleration effect on gas propagation compared to He I. While earlier studies suggested this distinction\cite{dhuleyGasPropagationLiquid2016}, their conclusions were constrained by experimental limitations described in Ref.~\cite{garceauHeatMassTransfer2019}. The present study offers more definitive evidence that the He II-cooled tube provides greater deceleration, attributed to the bath's superior heat transfer capabilities. These differences arise from the distinct thermal mechanisms at play: convective heat transfer in He I versus thermal counterflow in He II \cite{Van_Sciver-2012-HeCryo}. Fig.~\ref{fig:fig3}b illustrates He II rise times for four tank pressures (50, 100, 150, and 200 kPa). At 50 kPa, the lowest inlet mass flow rate resulted in the slowest gas propagation and most pronounced deceleration, while at 200 kPa, the gas reached a average velocity of approximately 10 m/s, with a comparatively mild deceleration effect.

\subsection{Theoretical modeling}\label{sec:model}
\begin{figure*}[!htb]
		\centering
		\includegraphics[scale=0.45]{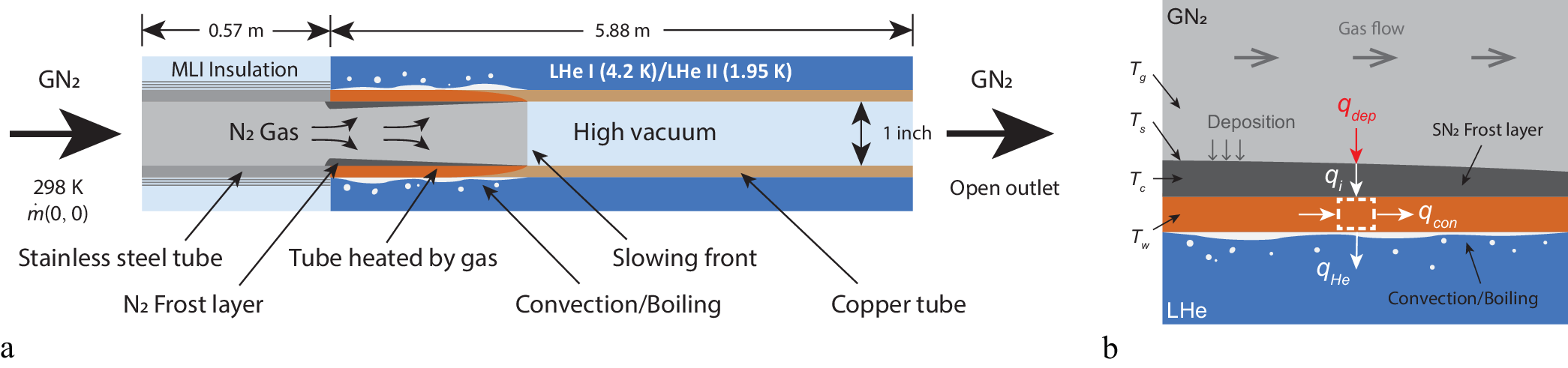}
		\caption{Schematic illustrating of (a) GN$_2$ propagation and deposition in a LHe-cooled vacuum tube and (b) radial heat transfer through the frost layer and tube wall, copied from \cite{baoHeatMassTransfer2020}.\label{fig:fig4} }
\end{figure*}

To interpret our experimental observations, a 1D theoretical model was developed, incorporating conservation equations for the propagating GN\(_2\), a phenomenological model for gas condensation on tube inner surface, and empirical correlations for heat transfer to He I and He II baths, as described in prior studies~\cite{baoHeatMassTransfer2020,garceauHeatMassTransfer2021,bao2024freeze}. This 1D approximation is justified by the tube's small diameter-to-length ratio. Fig.~\ref{fig:fig4}a schematically illustrates the computational domain. The gas inlet (\(x = 0\)) is defined as the location immediately downstream of the venturi tube. The system includes a 0.57-meter insulated stainless steel section, followed by a 5.88-meter LHe-cooled copper tube, designed to replicate the experimental setup. The GN\(_2\) velocity at the inlet is set to the speed of sound, as the flow is choked at the venturi. The experimentally measured mass flow rate at the inlet was incorporated into the numerical simulations~\cite{garceauHeatMassTransfer2019}. Additionally, the model includes tube dimensions and other relevant parameters that precisely match the experimental conditions outlined in Section~\ref{sec:exp_setup}, ensuring direct comparability with the experimental observations.

\subsubsection{Model on gas dynamics, condensation, and heat transfer}\label{sec:model_overall}
The propagation of GN\(_2\) gas in the tube and the evolution of its properties are governed by the conservation equations for N$_2$ mass, momentum, and energy, as outlined below:

\begin{equation}
\frac{\partial\rho_g}{\partial t}+\frac{\partial}{\partial x}(\rho_g v) = -\frac{4}{D_1} \dot{m}_{c},
\label{eq:modConsMass}
\end{equation}
\begin{equation}
				\frac{\partial}{\partial t}(\rho_g v) + \frac{\partial}{\partial x}(\rho_g v^2) = -\frac{\partial 	P}{\partial x} - \frac{4}{D_1}\dot{m}_{c} v,
				\label{eq:modConsMome}
\end{equation}
\begin{equation}
\begin{aligned}
\frac{\partial}{\partial t}\left[\rho_g\left(\varepsilon_g+ \frac{1}{2} v^2\right)\right]+\frac{\partial}{\partial x}\left[\rho_g v \left(\varepsilon_g+\frac{1}{2} v^2 + \frac{P}{\rho_g}\right)\right]=\\
-\frac{4}{D_1} \dot{m}_{c} \left(\varepsilon_g+\frac{1}{2}v^2 + \frac{P}{\rho_g} \right) - \frac{4}{D^2_1} Nu\cdot k_g (T_g-T_s).
\label{eq:modConsEner}
\end{aligned}
\end{equation}

The parameters used in the equations above are defined in the Nomenclature table. The Nusselt number Nu for GN$_2$ convective heat transfer is calculated using the Sieder-Tate correlation \cite{incroperaFundamentalsHeatMass2007}:
\begin{equation}
Nu=0.027Re^{4/5}Pr^{1/3}(\mu_\text{g}/\mu_\text{s})^{0.14}.
\label{eq:Nu}
\end{equation}
The model assumes the ideal-gas equation of state for N$_2$ gas:

\begin{equation}
P M_g =\rho_g R T_g.
\label{eq:IG}
\end{equation}
which is justified since the compressibility of N$_2$ remains close to unity throughout the experiment~\cite{baoHeatMassTransfer2020}. The terms containing $\dot{m}_c$ on the right-hand side of Eqs.~(\ref{eq:modConsMass})–(\ref{eq:modConsEner}) represent the effects of N$_2$ condensation on the inner tube surface. Here, $\dot{m}_c$ denotes the mass deposition rate per unit inner surface area of the tube, which is calculated using the Hertz-Knudsen relation with the Schrage modification~\cite{Collier-1994-ConvBoilCond}:
\begin{equation}
\dot{m}_{c} = \sqrt{ \frac{M_g}{2\pi R}}\left({\Gamma\sigma_c}\frac{P}{\sqrt{T_g}}-\sigma_e\frac{P_s}{\sqrt{T_s}} \right).
\label{eq:mc}
\end{equation}
In this context, $P_s$ represents the saturated vapor pressure at the frost-layer surface temperature $T_s$. The coefficients $\sigma_c$ and $\sigma_e$ are empirical condensation and evaporation coefficients, which are typically similar and close to unity for very cold surfaces~\cite{persad-2016_Chem.Rev.}. For our subsequent numerical simulations, we adopt a value of 0.95 for both coefficients. The Schrage parameter $\Gamma$ accounts for deviations from the Maxwellian velocity distribution of the molecules resulting from condensation on the wall. In terms of the mean flow velocity towards the cold wall, defined as $u = \dot{m}_{c}/\rho_g$, $\Gamma$ can be calculated as~\cite{Collier-1994-ConvBoilCond}:
\begin{equation}
\Gamma(\beta) = exp\left({-\beta^2}\right) +\beta\sqrt{\pi} \left[1+erf\left( {\beta}\right)\right].
\label{eq:gamma}
\end{equation}
where $\beta=u/u_T$, with $u_T=\sqrt{2RT_g/M_g}$ representing the thermal velocity of the gas molecules. It is important to note that since $\Gamma$ depends on $\dot{m}_c$ through $u$, Eq.~\ref{eq:mc} must be solved self-consistently at each time step to determine the evolution of $\dot{m}_c$. The frost-layer surface temperature, $T_s$, required for solving the above equations, can be determined by assessing the radial heat transfer processes across the frost layer, as shown in Fig.~\ref{fig:fig4}b:
\begin{equation}
	\rho_{SN} C_{SN} \delta \frac{\partial T_c}{\partial t} = \dot{m}_c(\frac{v^2}{2} + {\hat{h}}_g - {\hat{h}}_s) + \frac{1}{D_1}Nu\cdot k_g(T_g-T_s) - q_{con}.
	\label{eq:frost}
\end{equation}
where $T_c=(T_w+T_s)/2$ denotes the frost layer center temperature, and $q_{con}=k_{SN}(T_s-T_w)/\delta$ is the heat flux conducted through the frost layer thickness $\delta$, whose change rate is given by $\dot{\delta}=\dot{m}_c/\rho_{SN}$\,\cite{stephan1987-Physical-chemical-ref,scott1976-Physics_report}. The solid nitrogen properties are taken from Refs.~\cite{cook-1976_Cryogenicsa, scott1976-Physics_report}. The variation of the copper tube wall temperature $T_w$ is described by:
\begin{equation}
	\rho_w C_{w} \frac{D^2_2 - D^2_1}{4D_1} \frac{\partial T_w}{\partial t} = q_{con} - q_{He} \frac{D_2}{D_1}+\frac{D^2_2 - D^2_1}{4D_1} k_w \frac{\partial^2 T_w}{\partial x^2}.
	\label{eq:wall}
\end{equation}
In this equation, we neglect the temperature gradient across the copper tube thickness, which is reasonable given copper’s high thermal conductivity and the tube’s thin wall. The parameter $q_{He}$ in Eq.~\ref{eq:wall} denotes the instantaneous heat flux from the tube's outer surface to the LHe (He I or He II) bath, which is modeled separately and will be detailed in the following section. The physical properties of GN$_2$, copper, and stainless steel used in our simulations are sourced from the literature~\cite{stephan1987-Physical-chemical-ref, flynn2004cryogenic, lemmon2007nist, arp2005-INC}.

\subsubsection{LHe heat transfer modeling}

\begin{figure}[!tb]
	\centering
	\includegraphics[width=\columnwidth]{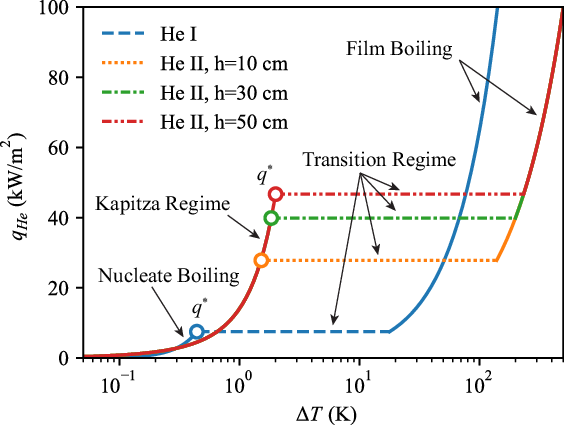}
	\caption{Representative correlation curves show the heat flux $ q_{He}$ from the tube outer surface to the LHe bath as a function of the temperature difference $\Delta T = T_w - T_b$, where $T_b$ is 4.2 K for He I and 1.9 K for He II, copied from \cite{bao2024freeze}.
	\label{fig:fig5}}
\end{figure}

Since temperature relaxation in LHe to a steady state occurs much faster than the slow gas propagation~\cite{van_sciver_helium_2012, Bao-2021-PRB, Sanavandi-2022-PRB}, we can simplify our model by applying steady-state heat transfer correlations between solid surfaces and LHe to calculate $q_{He}$~\cite{smith-1969_Cryogenicsa}. Fig.~\ref{fig:fig5} illustrates representative correlation curves for $q_{He}$, showing three distinct regimes based on the temperature difference $\Delta T = T_w - T_b$ between the tube wall and the LHe bath.

For He I, at small temperature differences $\Delta T_\text{w} < 0.1$ K, heat transfer near the wall is governed by natural convection with a coefficient of 0.375 kW/(m$^2\cdot$K)~\cite{lantz_heat_2007}. As $\Delta T_\text{w}$ rises, vapor bubbles form, moving the system into the nucleate boiling regime, which is described by the Kutateladze correlation\cite{van_sciver_helium_2012}. At a peak heat flux $q^*$ \cite{zuber_hydrodynamic_1961}, vapor bubbles coalesce into a film, triggering rapid wall temperature increases until film boiling stabilizes, at which point the Breen-Westwater correlation \cite{breen_effect_1962} applies:
\begin{equation}
			q_\text{He}=B_\text{w}\Delta T_\text{w}^{5/4}.
			\label{eq:bw}
\end{equation}
where $B_\text{w}$ is a constant that slightly depends on surface.

Different from He I, heat transfer in bulk He II occurs through a counterflow process, where the normal fluid moves heat away from the source while the superfluid flows oppositely to conserve mass~\cite{Van_Sciver-2012-HeCryo}. At heat fluxes beyond a critical value, quantized vortices form, disrupting this flow and increasing the temperature gradient~\cite{Vinen-1957-PRS-III}. Extensive studies on He II counterflow, both theoretical and experimental, have been conducted~\cite{Schwarz-1978-PRB, Martin-1983-PRB, Bewley-2008-PNAS, Chagovets-2011-PF, Mineda-2013-PRB, Svancara-2018-PRF}, with recent work in our lab covering both steady-state and transient behaviors~\cite{Marakov-2015-PRB, Gao-2016-JETP, Bao-2019_PRApplied, Bao-2021-PRB}. For heat transfer between solid surfaces and He II, two regimes exist: the Kapitza regime (low heat flux) and the film boiling regime (high heat flux)~\cite{Van_Sciver-2012-HeCryo, Murakami-2015-IOP}. In the Kapitza regime, heat flux $q_\text{He}$ depends on the wall and bath temperatures $T_w$ and $T_{bath}$ through:
\begin{equation}
q_\text{He}=\alpha\left(T^{n}_{w}-T^{n}_{b}\right).
\end{equation}
where $\alpha$ and $n$ vary by material and surface quality~\cite{Claudet-1981-ACE, Kashani-1985-Cryo}. For oxidized copper in He II, we use $\alpha = 0.5$ kW/m$^2$K$^n$ and $n = 3.5$~\cite{Kashani-1985-Cryo}. At the peak heat flux $q^*$, the system transitions from Kapitza to film boiling as a vapor film forms, causing a rapid rise in $\Delta T = T_w - T_b$. The peak flux $q^*$ is geometry- and pressure-dependent, and for a cylindrical heater of diameter $D_2$, it is approximated by~\cite{Van_Sciver-2012-HeCryo}:
\begin{equation}
q^{*}(T_{b},x)=\left( \frac{2\psi}{D_2/2}\int_{T_{b}}^{T'(x)}\frac{dT}{f(T)} \right)^{1/3}.
\label{Eq:peakHeat}
\end{equation}	
where $f(T)$ is the heat conductivity of He II, with $\psi$ as an empirical correction factor~\cite{VanSciver-1980-ACE}. In the film boiling regime, we assume $q_\text{He} = h_{film} (T_w - T_{bath})$, where $h_{film}$ is taken as 200 W/m$^2 \cdot$K based on available data for similar conditions~\cite{Goodling-1969-ACE, Bretts-1975-ACE}.
\subsubsection{Model validation and results}
We performed numerical simulations to solve the 1D gas dynamics and radial heat transfer equations using a two-step, first-order Godunov-type finite-difference method~\cite{danaila_introduction_2007}. The Godunov scheme is widely recognized for its efficiency and robustness in addressing nonlinear hyperbolic equations~\cite{sod_survey_1978}. In what follows, we present simulation results for comparison with experimental observations to validate the model. Additionally, we discuss key insights into the heat flux to the tube wall and helium bath, which are essential considerations for accelerator beamline applications~\cite{wisemanLossCavityVacuum1994, dalesandroExperimentTransientEffects2012}.

To achieve the best match between the simulated and observed gas dynamics, two parameters were adjusted: the coefficient $B_\text{w}$ in Eq.~\ref{eq:bw}, representing the Breen-Westwater correlation for He I film boiling heat flux, and the parameter $\psi$ in Eq.~\ref{Eq:peakHeat}, describing the integral correlation of $q^*$ for He II. The He I experiments at various inlet mass flow rates were accurately reproduced with an optimal $B_\text{w} = 0.021$  W/(cm$^2$·K$^{5/4}$). For He II experiments, where the immersion depth varies along the experimental tube, $\psi$ was found to depend on the inlet mass flow rate and ranged from 0.4 to 2~\cite{garceauHeatMassTransfer2021}. The effectiveness of these tuned parameters is illustrated in Fig.~\ref{fig:fig2}, where the solid lines represent the calculated tube-wall temperature profiles at sensor locations for representative runs: (a) He I at 100 kPa tank pressure and (b) He II at 150 kPa. The increasing separation between temperature curves over time agrees nicely with experimental observations. Further validation is provided in Fig.~\ref{fig:fig3}a, which compares the simulated GN\(_{2}\) arrival times, $t_r(x)$, for He I and He II-cooled tubes under the same inlet gas pressure. Figure~\ref{fig:fig3}b extends this comparison to He II experiments at different tank pressures. The rise-time curves show excellent agreement with experimental data across all runs, clearly demonstrating that the model successfully replicates the gas propagation dynamics for both He I and He II conditions.

\begin{figure}[!tb]
	\centering
	\includegraphics[width=\columnwidth]{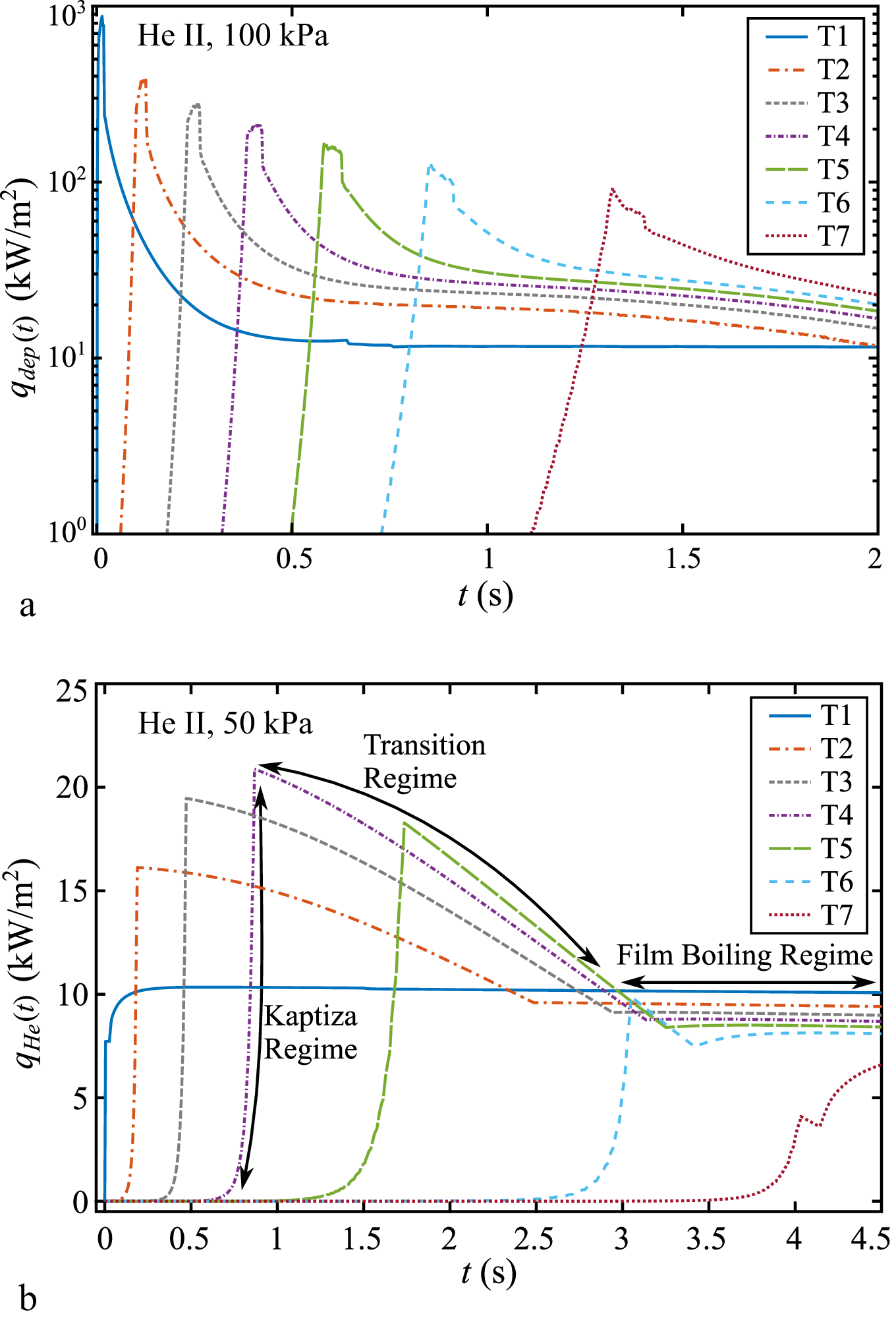}
	\caption{Simulated heat flux deposited to (a) the tube wall $(q_{dep})$ with a tank pressure of 100 kPa and (b) to the He II bath $(q_{He})$ with a tank pressure of 50 kPa, copied from \cite{garceauHeatMassTransfer2021}.
	\label{fig:fig6}}
\end{figure}

Building on the validated model, we performed an in-depth analysis of gas dynamics, frost growth, and heat deposition during gas propagation under a vacuum break scenario. Detailed findings can be found in our previous publications \cite{garceauHeatMassTransfer2021,baoHeatMassTransfer2020}. Here, we focus on a subset of the obtained heat deposition results, as these are experimentally challenging to measure precisely but can strongly impact accelerator performance. Specifically, we present the analysis of heat deposition to the He II bath as a representative case. To evaluate heat deposition to He II, we first calculated the total heat flux, $q_{dep}$, at sensor locations using the optimized model. Fig.~\ref{fig:fig6}a shows \( q_{dep}(t) \) curves for a representative run at 100 kPa tank pressure, revealing initial peak heat deposition rates exceeding $10^2$ kW/m$^2$. These heat deposition peaks correspond to the arrival of the gas front, where the mass deposition rate, \( \dot{m}_c \), and consequently $q_{dep}$, are highest as the gas first contacts the cold, clean wall surface. As frost accumulates on the wall, the wall temperature, $T_w$, rises sharply to around 50 K, leading to a rapid decline in $\dot{m}_c$ and $q_{dep}$. In the subsequent saturation phase, $T_w$ stabilizes, indicating an approximate equilibrium between $q_{dep}$ and the heat flux $q_{He}$ into the He II bath. Over time, $q_{dep}$ at sensor locations becomes primarily governed by the local $q_{He}$.

Next, we analyzed $q_{He}$, the heat flux into the He II bath. Fig.~\ref{fig:fig6}b illustrates $q_{He}$ over time at sensor locations for a tank pressure of 50 kPa. Initially, $q_{He}$ spikes as $T_w$ rises, reflecting the rapid transition of heat transfer from the Kapitza regime to the transition regime. In this regime, $q_{He}$ is limited by the peak heat flux, $q^*_{0}(T_b, x)$, which is higher at sensors submerged deeper in the He II bath. Following the peak, $q_{He}$ gradually declines due to the increasing bath temperature, \( T_b \) (see Fig.~\ref{fig:fig2}b), and levels off after approximately 3 seconds at most sensor locations. This plateau corresponds to the film boiling regime, where $q_{He}$ is regulated by the saturated wall temperature, $T_w$.

This model effectively captures the dynamics of heat flux deposition and dissipation to the helium bath, enabling accurate simulation of heat flux behavior under various operational conditions.
Moreover, our simulations reveal an intriguing phenomenon: at a given inlet mass flow rate, gas propagation nearly halts beyond a specific distance from the initial condensation point. We refer to this distance as the "freeze range". Understanding the freeze range could aid accelerator engineers in developing protocols to better manage frost-layer contamination within the beamline tube. Through a systematic numerical study, we derived a robust correlation for the freeze range, as detailed in Ref.~\cite{bao2024freeze}:
\begin{equation}
	x_F=a\cdot D_1^b\cdot\omega^c,
	\label{eq:fitmodel}
\end{equation}
where \( a \), \( b \), and \( c \) are empirical parameters. Table~\ref{tab:tab2} lists the optimal values of these parameters that render the best fits to the simulation data for both the He I and He II cases.

{\centering	
	\begin{table}[ht]
		\caption{Optimal parameters for evaluating freeze range of GN$_2$ propagation in He I- and He II-cooled copper tubes, copied from \cite{bao2024freeze}.\label{tab:tab2}}
		\centering
		\begin{tabular}{c c c c}	
			\hline
			Parameter		& He I		& He II ($h$=50 cm)& Unit			 		\\
			\hline
			$a$    			& 0.074		& 0.018				& m$^{2c-b+1} \cdot$kg$^{-c} \cdot$s$^{c}$	\\
			$b$         	& 0.915		& 1.023	 			& 1  		  			\\
			$c$      	 	& 1.084		& 1.395				& 1   		 			\\
			\hline	
		\end{tabular}
	\end{table}
}

\section{\label{sec:future}Recent progress and future perspectives}
In our earlier experiments and modeling of sudden vacuum loss in LHe-cooled tubes, we focused on a long tube with a uniform cross-section. However, real beamline tubes, which incorporate a series of superconducting radio frequency (SRF) cavities, exhibit highly non-uniform cross-sections. These SRF cavities, which are critical components of particle accelerators, are designed with complex geometries—often elliptical in shape—tailored to specific factors such as particle type and target energy levels~\cite{padamsee2015design,darve2014ess,dylla1994development,padamseerf,padamsee2008rf}. The larger surface area within these bulky SRF cavities provides an expanded condensing surface for incoming gas, which is expected to significantly influence both gas dynamics and heat deposition in LHe by enhancing condensation effects. To advance our understanding of how these cavities affect gas propagation and heat transfer during vacuum failures, we have planned experimental system modifications and additional modeling efforts.

\subsection{Modified setup and preliminary results}
\begin{figure}[!tb]
	\centering
	\includegraphics[width=\columnwidth]{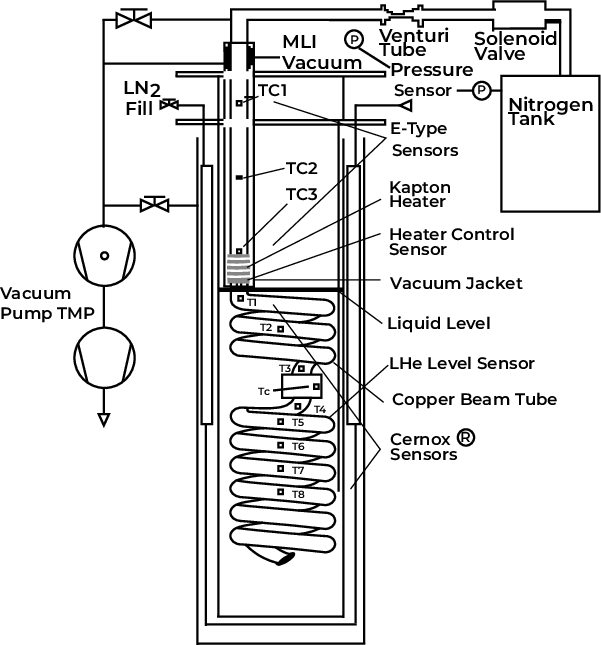}
	\caption{Schematic of the new vacuum break system with an inline cylindrical
cavity.
	\label{fig:fig7}}
\end{figure}

The experimental setup has been modified to include a new helical copper tube with a single inserted cavity, positioned between the third and fourth turns, as illustrated in Fig.~\ref{fig:fig7}. For simplicity in analysis and fabrication, the cavity was designed with a cylindrical shape. To approximate the dimensions of an actual accelerator system, the cavity was scaled using an aspect ratio of 1:3:1.8 for the inlet diameter, maximum diameter, and cavity length, based on representative values for SRF cavities from the literature \cite{padamsee2015design}. The cavity wall has a thickness of 1.25 mm. The copper tube has an inner diameter of 25.4 mm and a wall thickness of 1.25 mm, while the coil's diameter and pitch have been adjusted to 200 mm and 50 mm, respectively, to accommodate the inserted cavity.

As shown in Fig.~\ref{fig:fig7}, nine Lake Shore Cernox\textsuperscript{\textregistered} sensors were installed to measure surface temperatures along the copper tube. Sensor \(T_3\) is positioned just upstream of the cavity, while \(T_4\) is located immediately downstream. Additionally, sensor \(T_c\) is mounted on the side wall of the cavity. A preliminary experiment was conducted using He I at a buffer tank pressure of 100 kPa~\cite{garceau2022vacuum,garceau2022heat}, following the procedure outlined in Sec.~\ref{sec:exp_procedure}. The results revealed an intriguing feature: the temperature at sensor \(T_4\) (located just downstream of the cavity) begins to rise earlier than at \(T_c\) (on the cavity side wall), despite \(T_4\) being downstream of \(T_c\). This observation suggests that the gas propagates through the cavity before completely filling it. Moving forward, systematic experiments will be conducted in both He I and He II baths, with variations in the position, size, and shape of the cavity, as well as configurations involving multiple cavities. This effort aims to develop a comprehensive understanding of how chains of bulky cavities influence gas propagation dynamics in accelerator beamlines during sudden vacuum loss events.

\subsection{2D model}
The anisotropic GN$_2$ flow in the bulky cavity in our preliminary experiment suggests that the 1D simulation model will be insufficient to accurately capture the gas dynamics in this complex geometry. To address this limitation, we plan to extend the existing 1D model as detailed in Sec.~\ref{sec:model} into a more comprehensive two-dimensional (2D) model that better represents gas propagation within the cavity.

\begin{figure*}[!htb]
	\centering
	\includegraphics[scale=0.36]{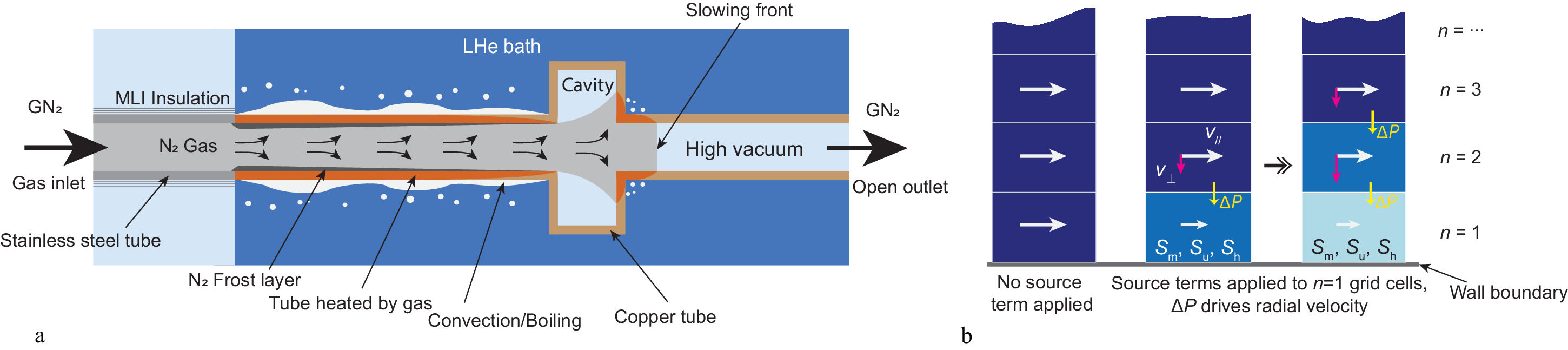}
	\caption{Schematic illustrating of (a) GN$_2$ propagation and deposition in a LHe-cooled vacuum tube with an inline cylindrical
cavity and (b) mass transfer calculation for the 2D model.
	\label{fig:fig8}}
\end{figure*}

The computational domain of the 2D model for nitrogen gas propagation in a LHe-cooled non-uniform tube following a vacuum break is illustrated in Fig.~\ref{fig:fig8}a. This domain replicates the geometry of the modified experimental tube, including the inserted cylindrical cavity. The gas flow dynamics within the tube are governed by the conservation equations for mass, streamwise and radial momentum, and energy. Turbulent effects are incorporated into the model to ensure an accurate representation of the flow characteristics:
\begin{equation}
\frac{\partial \rho_g }{\partial t} + \nabla \cdot (\rho_g  \vec{v}) = S_m,
\label{Eq:2D_mass}
\end{equation}	
\begin{equation}
\frac{\partial}{\partial t} (\rho_g  \vec{v}) + \nabla \cdot (\rho_g  \vec{v} \vec{v}) = -\nabla P + \nabla \cdot \bar{\bar{\tau}} + S_u,
\label{Eq:2D_momentum}
\end{equation}	
\begin{equation}
\frac{\partial}{\partial t} (\rho_g  E) + \nabla \cdot \left[\vec{v} (\rho_g E + P)\right] =\nabla \cdot  (k_{\text{eff}} \nabla T + \tau_{\text{eff}} \cdot \vec{v}) + S_h.
\label{Eq:2D_energy}
\end{equation}
where \(\bar{\bar{\tau}}\) and \(E\) denote the stress tensor and the specific internal energy of nitrogen gas, respectively. The effective thermal conductivity is expressed as \(k_{\text{eff}} = k + k_t\), where \(k_t\) is the turbulent thermal conductivity and is computed using the standard \(k\)-\(\varepsilon\) turbulence model \cite{launder1972lectures}. The effective stress tensor is defined as \(\tau_{\text{eff}} = \frac{\mu_{\text{eff}}}{\mu} \bar{\bar{\tau}}\), with \(\mu_{\text{eff}}\) representing the effective viscosity, given by \(\mu_{\text{eff}} = \frac{k_{\text{eff}} \text{Pr}}{c_p}\).

The source terms ($S_x$) on the right-hand side of Eqs.~(\ref{Eq:2D_mass})–(\ref{Eq:2D_energy}) involve the mass deposition rate $\dot{m}_c$ of the GN$_2$ on the tube inner wall. In the 1D model, the tube is divided into \(N\) grid points along its length \(L\), with each point representing a tube segment of length \(L/N\). Source terms related to mass deposition on the copper tube wall are applied to the whole segment, influencing the average properties of the gas within. In contrast, the 2D model discretizes the tube cross-section into a two-dimensional computational mesh. The $\dot{m}_c$-related source terms are introduced to account for condensation effects and are applied exclusively to grid cells adjacent to the tube wall. Fig.~\ref{fig:fig8}b illustrates this specific implementation, highlighting that these terms are restricted to the \(n=1\) grid cells adjacent to the wall. The detailed expressions for these source terms are provided in Table~\ref{tab:source terms}, where \(\delta_g\) represents the height of the corresponding grid cell. Iterative calculations of
$\dot{m}_c$ follow Eqs.~(\ref{eq:mc})–(\ref{eq:gamma}), consistent with the 1D model.

The nitrogen condensation-driven mass transfer alters gas-phase characteristics in \(n=1\) grid cells, which in turn, propagates changes throughout the computational domain. Fig.~\ref{fig:fig8}b uses grid cell shading to indicate nitrogen gas density. As shown, the application of source terms in the \(n=1\) cells reduces local nitrogen gas density, creating a density—and consequently pressure—gradient between \(n=1\) and \(n=2\) cells. This gradient induces a radial velocity (\(v_\perp\)) in the \(n=2\) cells, further reducing their density. The resulting pressure difference between \(n=2\) and \(n=3\) cells drives \(v_\perp\) in the \(n=3\) cells. This cascade of radial pressure differences progressively drives radial flow across the domain, which transports the streamwise momentum towards the tube wall, effectively accounting for both the gas condensation and the gradual slowing of gas propagation. Radial heat transfer is calculated similarly to the 1D model.

{\centering	
	\begin{table}[ht]
		\caption{Source terms in the governing equations for the 2D model.\label{tab:source terms}}
		\centering
		\begin{tabular}{c c c }	
			\hline
			& $n$=1& $n>$1\\
			\hline
			$S_m$& $-\dot{m}_c/\delta_g$& 0\\
			$S_u$& $-\dot{m}_c\vec{v}/\delta_g$& 0\\
			$S_h$& $-\dot{m}_cE/\delta_g$& 0\\
			\hline	
		\end{tabular}
	\end{table}
}

Unlike the 1D model, which provides only an average view of gas dynamics along the tube, the proposed 2D model captures the detailed distribution of nitrogen gas properties and velocity fields across the tube’s cross-section, allowing for direct comparisons with experiments involving inserted bulky cavities. By comparing the simulation results with data from upcoming experiments using the same tube geometry and initial conditions, we will refine the model to develop a robust tool for accurately describing heat and mass transfer processes in non-uniform beamline tubes.

\section{\label{sec:sum}Summary}
In this review, we have outlined the progress made in our cryogenics lab in understanding the complex heat and mass transfer dynamics associated with sudden vacuum breaks in LHe-cooled tube systems. Systematic experimental studies in uniform copper tubes cooled by both He I and He II have provided significant insights into nitrogen gas propagation, condensation, and frost formation. These observations are well captured by a 1D theoretical model that accounts for the gas dynamics, mass deposition, and heat transfer to helium, forming a strong foundation for this field. Nevertheless, further work is necessary to address the complexities of real-world geometries. As revealed in our more recent experiments that incorporated a bulky cavity into the uniform tube system—mimicking the geometry of SRF cavities in real accelerator beamlines—, there is a strongly anisotropic gas flow within the cavity. These findings highlight the need for additional research into the behavior of gas dynamics and heat transfer in non-uniform geometries. Future efforts should include systematic experimental studies involving tube systems with multiple inserted cavities of varying shapes and configurations to better understand these phenomena. In parallel, the development of the 2D theoretical models will enable more accurate simulations of gas dynamics in these complex systems. Advancing this understanding will provide critical insights for designing safer and more efficient accelerator beamlines, aiding engineers in developing robust safety protocols for vacuum failure events.

\section*{Acknowledgments}
This work is supported by U.S. Department of Energy, United States under Grant No. DE-SC0020113. The experiment was conducted at the National High Magnetic Field Laboratory, United States, which is supported by National Science Foundation, United States Cooperative Agreement No. DMR-2128556 and the State of Florida.







\bibliographystyle{elsarticle-num-names}
\bibliography{Ref}

\begin{thebibliography}{79}
\expandafter\ifx\csname natexlab\endcsname\relax\def\natexlab#1{#1}\fi
\providecommand{\url}[1]{\texttt{#1}}
\providecommand{\href}[2]{#2}
\providecommand{\path}[1]{#1}
\providecommand{\DOIprefix}{doi:}
\providecommand{\ArXivprefix}{arXiv:}
\providecommand{\URLprefix}{URL: }
\providecommand{\Pubmedprefix}{pmid:}
\providecommand{\doi}[1]{\href{http://dx.doi.org/#1}{\path{#1}}}
\providecommand{\Pubmed}[1]{\href{pmid:#1}{\path{#1}}}
\providecommand{\bibinfo}[2]{#2}
\ifx\xfnm\relax \def\xfnm[#1]{\unskip,\space#1}\fi
\bibitem[{Padamsee(2015)}]{padamsee2015design}
\bibinfo{author}{H.~Padamsee},
\newblock \bibinfo{title}{Design topics for superconducting {RF} cavities and ancillaries},
\newblock \bibinfo{journal}{arXiv preprint arXiv:1501.07129}  (\bibinfo{year}{2015}).
\bibitem[{Harrison(2002)}]{harrisonLossVacuumExperiments2002}
\bibinfo{author}{S.~M. Harrison},
\newblock \bibinfo{title}{Loss of vacuum experiments on a superfluid helium vessel},
\newblock \bibinfo{journal}{IEEE Transactions on Applied Superconductivity} \bibinfo{volume}{12} (\bibinfo{year}{2002}) \bibinfo{pages}{1343--1346}. \DOIprefix\doi{https://doi.org/10.1109/TASC.2002.1018651}.
\bibitem[{Petitpas and Aceves(2013)}]{petitpasModelingSuddenHydrogen2013}
\bibinfo{author}{G.~Petitpas}, \bibinfo{author}{S.~M. Aceves},
\newblock \bibinfo{title}{Modeling of sudden hydrogen expansion from cryogenic pressure vessel failure},
\newblock \bibinfo{journal}{International Journal of Hydrogen Energy} \bibinfo{volume}{38} (\bibinfo{year}{2013}) \bibinfo{pages}{8190--8198}. \DOIprefix\doi{https://doi.org/10.1016/j.ijhydene.2012.03.166}.
\bibitem[{Heidt et~al.(2014)Heidt, Grohmann, and S{\"u}{\ss}er}]{heidtModelingPressureIncrease2014}
\bibinfo{author}{C.~Heidt}, \bibinfo{author}{S.~Grohmann}, \bibinfo{author}{M.~S{\"u}{\ss}er},
\newblock \bibinfo{title}{Modeling the pressure increase in liquid helium cryostats after failure of the insulating vacuum},
\newblock in: \bibinfo{booktitle}{AIP Conference Proceedings}, volume \bibinfo{volume}{1573}, \bibinfo{publisher}{American Institute of Physics}, \bibinfo{year}{2014}, pp. \bibinfo{pages}{1574--1580}.
\bibitem[{Xie et~al.(2010)Xie, Li, and Wang}]{xieStudyHeatTransfer2010}
\bibinfo{author}{{\relax GF}.~Xie}, \bibinfo{author}{{\relax XD}.~Li}, \bibinfo{author}{{\relax RS}.~Wang},
\newblock \bibinfo{title}{Study on the heat transfer of high-vacuum-multilayer-insulation tank after sudden, catastrophic loss of insulating vacuum},
\newblock \bibinfo{journal}{Cryogenics} \bibinfo{volume}{50} (\bibinfo{year}{2010}) \bibinfo{pages}{682--687}. \DOIprefix\doi{https://doi.org/10.1016/j.cryogenics.2010.06.020}.
\bibitem[{Ady et~al.(2014)Ady, Ziemianski, Vandoni, Kersevan, and Hermann}]{adyLeakPropagationDynamics2014}
\bibinfo{author}{M.~Ady}, \bibinfo{author}{D.~Ziemianski}, \bibinfo{author}{G.~Vandoni}, \bibinfo{author}{R.~Kersevan}, \bibinfo{author}{M.~Hermann}, \bibinfo{title}{Leak propagation dynamics for the {HIE-ISOLDE} superconducting linac}, \bibinfo{type}{Technical Report}, CERN, \bibinfo{year}{2014}.
\bibitem[{Bajko et~al.(2009)Bajko, Bertinelli, and {Catalan-Lasheras et al.}}]{bajkoReportTaskForce2009}
\bibinfo{author}{M.~Bajko}, \bibinfo{author}{F.~Bertinelli}, \bibinfo{author}{{Catalan-Lasheras et al.}}, \bibinfo{title}{Report of the task force on the incident of 19th September 2008 at the {LHC}}, \bibinfo{type}{Technical Report}, CERN, \bibinfo{address}{Geneva}, \bibinfo{year}{2009}.
\bibitem[{Seidel et~al.(2002)Seidel, Trines, and Zapfe}]{seidelFailureAnalysisBeam2002}
\bibinfo{author}{M.~Seidel}, \bibinfo{author}{D.~Trines}, \bibinfo{author}{K.~Zapfe}, \bibinfo{title}{Failure analysis of the beam vacuum in the superconducting cavities of the {TESLA} main linear accelerator}, \bibinfo{type}{Technical Report}, Dt. Elektronen-Synchrotron DESY, MHF-SL Group, \bibinfo{year}{2002}.
\bibitem[{Wiseman et~al.(1994)Wiseman, Crawford, Drury, Jordan, Preble, Saulter, and Schneider}]{wisemanLossCavityVacuum1994}
\bibinfo{author}{M.~Wiseman}, \bibinfo{author}{K.~Crawford}, \bibinfo{author}{M.~Drury}, \bibinfo{author}{K.~Jordan}, \bibinfo{author}{J.~Preble}, \bibinfo{author}{Q.~Saulter}, \bibinfo{author}{W.~Schneider},
\newblock \bibinfo{title}{Loss of cavity vacuum experiment at {CEBAF}},
\newblock \bibinfo{journal}{Advances in cryogenic engineering}  (\bibinfo{year}{1994}) \bibinfo{pages}{997--1003}.
\bibitem[{Wiseman et~al.(1991)Wiseman, Bundy, Kelley, and Schneider}]{wiseman1991cebaf}
\bibinfo{author}{M.~Wiseman}, \bibinfo{author}{R.~Bundy}, \bibinfo{author}{J.~P. Kelley}, \bibinfo{author}{W.~Schneider},
\newblock \bibinfo{title}{{CEBAF} cryounit loss of vacuum experiment},
\newblock \bibinfo{journal}{Applications of Cryogenic Technology}  (\bibinfo{year}{1991}) \bibinfo{pages}{287--303}. \DOIprefix\doi{https://doi.org/10.1007/978-1-4757-9232-4$\_$24}.
\bibitem[{Toro(2013)}]{toroRiemannSolversNumerical2013}
\bibinfo{author}{E.~F. Toro}, \bibinfo{title}{Riemann solvers and numerical methods for fluid dynamics: A practical introduction}, \bibinfo{publisher}{Springer Science \& Business Media}, \bibinfo{year}{2013}.
\bibitem[{Shapiro(1953)}]{shapiroDynamicsThermodynamicsCompressible1953}
\bibinfo{author}{A.~H. Shapiro},
\newblock \bibinfo{title}{The dynamics and thermodynamics of compressible fluid flow},
\newblock \bibinfo{journal}{New York: Ronald Press} \bibinfo{volume}{2} (\bibinfo{year}{1953}).
\bibitem[{Bosque(2014)}]{bosqueTransientHeatTransfer2014}
\bibinfo{author}{E.~Bosque}, \bibinfo{title}{Transient heat transfer to helium {II} due to a sudden loss of insulating vacuum}, Ph.D. thesis, The Florida State University, \bibinfo{year}{2014}.
\bibitem[{Welch(2001)}]{welchCapturePumpingTechnology2001}
\bibinfo{author}{K.~Welch}, \bibinfo{title}{Capture pumping technology}, \bibinfo{publisher}{Elsevier}, \bibinfo{year}{2001}.
\bibitem[{Dawson and Haygood(1965)}]{dawsonCryopumping1965}
\bibinfo{author}{{\relax JP}.~Dawson}, \bibinfo{author}{{\relax JD}.~Haygood},
\newblock \bibinfo{title}{Cryopumping},
\newblock \bibinfo{journal}{Cryogenics} \bibinfo{volume}{5} (\bibinfo{year}{1965}) \bibinfo{pages}{57--67}.
\bibitem[{Rogers(1966)}]{rogersExperimentalInvestigationsSolid1966}
\bibinfo{author}{{\relax KW}.~Rogers}, \bibinfo{title}{Experimental investigations of solid nitrogen formed by cryopumping}, \bibinfo{type}{Technical Report}, NATIONAL AERONAUTICS AND SPACE ADMINISTRATION, \bibinfo{year}{1966}.
\bibitem[{Tantos et~al.(2016)Tantos, Naris, and Valougeorgis}]{tantosGasFlowAdsorbing2016}
\bibinfo{author}{C.~Tantos}, \bibinfo{author}{S.~Naris}, \bibinfo{author}{D.~Valougeorgis},
\newblock \bibinfo{title}{Gas flow towards an adsorbing planar wall subject to partial gas-surface thermal accommodation},
\newblock \bibinfo{journal}{Vacuum} \bibinfo{volume}{125} (\bibinfo{year}{2016}) \bibinfo{pages}{65--74}. \DOIprefix\doi{https://doi.org/10.1016/j.vacuum.2015.12.002}.
\bibitem[{Brown et~al.(1970)Brown, Trayer, and Busby}]{brownCondensation3002500Gases1970}
\bibinfo{author}{{\relax RF}.~Brown}, \bibinfo{author}{{\relax DM}.~Trayer}, \bibinfo{author}{{\relax MR}.~Busby},
\newblock \bibinfo{title}{Condensation of 300--2500 {K} gases on surfaces at cryogenic temperatures},
\newblock \bibinfo{journal}{Journal of Vacuum Science and Technology} \bibinfo{volume}{7} (\bibinfo{year}{1970}) \bibinfo{pages}{241--246}. \DOIprefix\doi{https://doi.org/10.1116/1.1315808}.
\bibitem[{Boeckmann et~al.(2008)Boeckmann, Hoppe, Jensch, Lange, Maschmann, Petersen, and Schnautz}]{boeckmannExperimentalTestsFault2008}
\bibinfo{author}{T.~Boeckmann}, \bibinfo{author}{D.~Hoppe}, \bibinfo{author}{K.~Jensch}, \bibinfo{author}{R.~Lange}, \bibinfo{author}{W.~Maschmann}, \bibinfo{author}{B.~Petersen}, \bibinfo{author}{T.~Schnautz},
\newblock \bibinfo{title}{Experimental tests of fault conditions during the cryogenic operation of a {XFEL} prototype cryomodule},
\newblock in: \bibinfo{booktitle}{Proceedings of International Cryogenic Engineering Conference}, volume~\bibinfo{volume}{22}, \bibinfo{publisher}{HM Chang}, \bibinfo{year}{2008}, pp. \bibinfo{pages}{723--728}.
\bibitem[{Dalesandro et~al.(2014)Dalesandro, Dhuley, Theilacker, and Van~Sciver}]{dalesandroResultsSuddenLoss2014}
\bibinfo{author}{A.~A. Dalesandro}, \bibinfo{author}{R.~C. Dhuley}, \bibinfo{author}{J.~C. Theilacker}, \bibinfo{author}{S.~W. Van~Sciver},
\newblock \bibinfo{title}{Results from sudden loss of vacuum on scaled superconducting radio frequency cryomodule experiment},
\newblock in: \bibinfo{booktitle}{AIP Conference Proceedings}, volume \bibinfo{volume}{1573}, \bibinfo{publisher}{American Institute of Physics}, \bibinfo{year}{2014}, pp. \bibinfo{pages}{1822--1828}.
\bibitem[{Dhuley and Van~Sciver(2016{\natexlab{a}})}]{dhuleyPropagationNitrogenGas2016}
\bibinfo{author}{{\relax RC}.~Dhuley}, \bibinfo{author}{{\relax SW}.~Van~Sciver},
\newblock \bibinfo{title}{Propagation of nitrogen gas in a liquid helium cooled vacuum tube following sudden vacuum loss--part {II}: Analysis of the propagation speed},
\newblock \bibinfo{journal}{International Journal of Heat and Mass Transfer} \bibinfo{volume}{98} (\bibinfo{year}{2016}{\natexlab{a}}) \bibinfo{pages}{728--737}. \DOIprefix\doi{https://doi.org/10.1016/j.ijheatmasstransfer.2016.03.077}.
\bibitem[{Dhuley and Van~Sciver(2016{\natexlab{b}})}]{dhuleyPropagationNitrogenGas2016a}
\bibinfo{author}{{\relax RC}.~Dhuley}, \bibinfo{author}{{\relax SW}.~Van~Sciver},
\newblock \bibinfo{title}{Propagation of nitrogen gas in a liquid helium cooled vacuum tube following sudden vacuum loss--part {I}: Experimental investigations and analytical modeling},
\newblock \bibinfo{journal}{International Journal of Heat and Mass Transfer} \bibinfo{volume}{96} (\bibinfo{year}{2016}{\natexlab{b}}) \bibinfo{pages}{573--581}. \DOIprefix\doi{https://doi.org/10.1016/j.ijheatmasstransfer.2016.01.077}.
\bibitem[{Dhuley(2016)}]{dhuleyGasPropagationLiquid2016}
\bibinfo{author}{R.~C. Dhuley}, \bibinfo{title}{Gas propagation in a liquid helium cooled vacuum tube following a sudden vacuum loss}, Ph.D. thesis, The Florida State University, \bibinfo{year}{2016}.
\bibitem[{Garceau et~al.(2019{\natexlab{a}})Garceau, Bao, Guo, and Van~Sciver}]{garceauDesignTestingLiquid2019}
\bibinfo{author}{N.~Garceau}, \bibinfo{author}{S.~Bao}, \bibinfo{author}{W.~Guo}, \bibinfo{author}{{\relax SW}.~Van~Sciver},
\newblock \bibinfo{title}{The design and testing of a liquid helium cooled tube system for simulating sudden vacuum loss in particle accelerators},
\newblock \bibinfo{journal}{Cryogenics} \bibinfo{volume}{100} (\bibinfo{year}{2019}{\natexlab{a}}) \bibinfo{pages}{92--96}. \DOIprefix\doi{https://doi.org/10.1016/j.cryogenics.2019.04.012}.
\bibitem[{Garceau et~al.(2019{\natexlab{b}})Garceau, Bao, and Guo}]{garceauHeatMassTransfer2019}
\bibinfo{author}{N.~Garceau}, \bibinfo{author}{S.~Bao}, \bibinfo{author}{W.~Guo},
\newblock \bibinfo{title}{Heat and mass transfer during a sudden loss of vacuum in a liquid helium cooled tube--part {I}: Interpretation of experimental observations},
\newblock \bibinfo{journal}{International Journal of Heat and Mass Transfer} \bibinfo{volume}{129} (\bibinfo{year}{2019}{\natexlab{b}}) \bibinfo{pages}{1144--1150}. \DOIprefix\doi{https://doi.org/10.1016/j.ijheatmasstransfer.2018.10.053}.
\bibitem[{Garceau et~al.(2020)Garceau, Bao, and Guo}]{garceauEffectMassFlow2020}
\bibinfo{author}{N.~Garceau}, \bibinfo{author}{S.~Bao}, \bibinfo{author}{W.~Guo},
\newblock \bibinfo{title}{Effect of mass flow rate on gas propagation after vacuum break in a liquid helium cooled tube.},
\newblock in: \bibinfo{booktitle}{IOP Conference Series: Materials Science and Engineering}, volume \bibinfo{volume}{755}, \bibinfo{publisher}{IOP Publishing}, \bibinfo{year}{2020}, p. \bibinfo{pages}{012112}.
\bibitem[{Bao et~al.(2020)Bao, Garceau, and Guo}]{baoHeatMassTransfer2020}
\bibinfo{author}{S.~Bao}, \bibinfo{author}{N.~Garceau}, \bibinfo{author}{W.~Guo},
\newblock \bibinfo{title}{Heat and mass transfer during a sudden loss of vacuum in a liquid helium cooled tube--part {II}: Theoretical modeling},
\newblock \bibinfo{journal}{International Journal of Heat and Mass Transfer} \bibinfo{volume}{146} (\bibinfo{year}{2020}) \bibinfo{pages}{118883}. \DOIprefix\doi{https://doi.org/10.1016/j.ijheatmasstransfer.2019.118883}.
\bibitem[{Garceau et~al.(2021)Garceau, Bao, and Guo}]{garceauHeatMassTransfer2021}
\bibinfo{author}{N.~Garceau}, \bibinfo{author}{S.~Bao}, \bibinfo{author}{W.~Guo},
\newblock \bibinfo{title}{Heat and mass transfer during a sudden loss of vacuum in a liquid helium cooled tube-part {III}: Heat deposition in he {II}},
\newblock \bibinfo{journal}{International Journal of Heat and Mass Transfer} \bibinfo{volume}{181} (\bibinfo{year}{2021}) \bibinfo{pages}{121885}. \DOIprefix\doi{https://doi.org/10.1016/j.ijheatmasstransfer.2021.121885}.
\bibitem[{Garceau et~al.(2022)Garceau, Bao, and Guo}]{garceau2022vacuum}
\bibinfo{author}{N.~Garceau}, \bibinfo{author}{S.~Bao}, \bibinfo{author}{W.~Guo},
\newblock \bibinfo{title}{Vacuum break in a helium cooled tube with an inserted cavity},
\newblock in: \bibinfo{booktitle}{International Cryogenic Engineering Conference and International Cryogenic Materials Conference}, \bibinfo{organization}{Springer}, \bibinfo{year}{2022}, pp. \bibinfo{pages}{413--419}.
\bibitem[{Dalesandro et~al.(2012)Dalesandro, Theilacker, and Van~Sciver}]{dalesandro2012experiment}
\bibinfo{author}{A.~A. Dalesandro}, \bibinfo{author}{J.~Theilacker}, \bibinfo{author}{S.~Van~Sciver},
\newblock \bibinfo{title}{Experiment for transient effects of sudden catastrophic loss of vacuum on a scaled superconducting radio frequency cryomodule},
\newblock in: \bibinfo{booktitle}{AIP Conference Proceedings}, volume \bibinfo{volume}{1434}, \bibinfo{organization}{American Institute of Physics}, \bibinfo{year}{2012}, pp. \bibinfo{pages}{1567--1574}. \DOIprefix\doi{https://doi.org/10.1063/1.4707087}.
\bibitem[{Dalesandro et~al.(2014)Dalesandro, Dhuley, Theilacker, and Van~Sciver}]{dalesandro2014results}
\bibinfo{author}{A.~A. Dalesandro}, \bibinfo{author}{R.~C. Dhuley}, \bibinfo{author}{J.~C. Theilacker}, \bibinfo{author}{S.~W. Van~Sciver},
\newblock \bibinfo{title}{Results from sudden loss of vacuum on scaled superconducting radio frequency cryomodule experiment},
\newblock in: \bibinfo{booktitle}{AIP Conference Proceedings}, volume \bibinfo{volume}{1573}, \bibinfo{organization}{American Institute of Physics}, \bibinfo{year}{2014}, pp. \bibinfo{pages}{1822--1828}. \DOIprefix\doi{https://doi.org/10.1063/1.4860929}.
\bibitem[{Dalesandro et~al.(2012)Dalesandro, Theilacker, and Van~Sciver}]{dalesandroExperimentTransientEffects2012}
\bibinfo{author}{A.~A. Dalesandro}, \bibinfo{author}{J.~Theilacker}, \bibinfo{author}{S.~Van~Sciver},
\newblock \bibinfo{title}{Experiment for transient effects of sudden catastrophic loss of vacuum on a scaled superconducting radio frequency cryomodule},
\newblock in: \bibinfo{booktitle}{AIP Conference Proceedings}, volume \bibinfo{volume}{1434}, \bibinfo{publisher}{American Institute of Physics}, \bibinfo{year}{2012}, pp. \bibinfo{pages}{1567--1574}.
\bibitem[{Dhuley and Van~Sciver(2014)}]{dhuleySuddenVacuumLoss2014}
\bibinfo{author}{R.~C. Dhuley}, \bibinfo{author}{S.~W. Van~Sciver},
\newblock \bibinfo{title}{Sudden vacuum loss in long liquid helium cooled tubes},
\newblock \bibinfo{journal}{IEEE Transactions on Applied Superconductivity} \bibinfo{volume}{25} (\bibinfo{year}{2014}) \bibinfo{pages}{1--5}.
\bibitem[{Dhuley et~al.(2014)Dhuley, Bosque, and Van~Sciver}]{dhuley2014cryodeposition}
\bibinfo{author}{R.~Dhuley}, \bibinfo{author}{E.~Bosque}, \bibinfo{author}{S.~Van~Sciver},
\newblock \bibinfo{title}{Cryodeposition of nitrogen gas on a surface cooled by helium {II}},
\newblock in: \bibinfo{booktitle}{AIP Conference Proceedings}, volume \bibinfo{volume}{1573}, \bibinfo{organization}{American Institute of Physics}, \bibinfo{year}{2014}, pp. \bibinfo{pages}{626--632}. \DOIprefix\doi{https://doi.org/10.1063/1.4860760}.
\bibitem[{Dhuley and Van~Sciver(2015)}]{dhuley2015heat}
\bibinfo{author}{R.~Dhuley}, \bibinfo{author}{S.~Van~Sciver},
\newblock \bibinfo{title}{Heat transfer in a liquid helium cooled vacuum tube following sudden vacuum loss},
\newblock in: \bibinfo{booktitle}{IOP Conference Series: Materials Science and Engineering}, volume \bibinfo{volume}{101}, \bibinfo{organization}{IOP Publishing}, \bibinfo{year}{2015}, p. \bibinfo{pages}{012006}. \DOIprefix\doi{https://doi.org/10.1088/1757-899X/101/1/012006}.
\bibitem[{{National Institute of Standards and Technology}(2019)}]{nist_nitrogen_2019}
\bibinfo{author}{{National Institute of Standards and Technology}}, \bibinfo{title}{Thermodynamics source database: Nitrogen in {NIST} chemistry webbook, {SRD} 69}, \bibinfo{howpublished}{\url{https://webbook.nist.gov/chemistry}}, \bibinfo{year}{2019}. \bibinfo{note}{Web accessed Feb 2019}.
\bibitem[{Dhuley and Van~Sciver(2016)}]{dhuley2016epoxy}
\bibinfo{author}{R.~Dhuley}, \bibinfo{author}{S.~Van~Sciver},
\newblock \bibinfo{title}{Epoxy encapsulation of the {Cernox™ SD} thermometer for measuring the temperature of surfaces in liquid helium},
\newblock \bibinfo{journal}{Cryogenics} \bibinfo{volume}{77} (\bibinfo{year}{2016}) \bibinfo{pages}{49--52}. \DOIprefix\doi{https://doi.org/10.1016/j.cryogenics.2016.05.001}.
\bibitem[{Van~Sciver(2012)}]{Van_Sciver-2012-HeCryo}
\bibinfo{author}{S.~W. Van~Sciver}, \bibinfo{title}{Helium Cryogenics}, International cryogenics monograph series, \bibinfo{edition}{2nd} ed., \bibinfo{publisher}{Springer}, \bibinfo{address}{New York}, \bibinfo{year}{2012}.
\bibitem[{Bao et~al.(2024)Bao, Tang, Rababah, and Guo}]{bao2024freeze}
\bibinfo{author}{S.~Bao}, \bibinfo{author}{Y.~Tang}, \bibinfo{author}{Q.~Rababah}, \bibinfo{author}{W.~Guo},
\newblock \bibinfo{title}{Freeze range of a condensing gas propagating in a liquid helium-cooled tube},
\newblock \bibinfo{journal}{Thermal Science and Engineering Progress} \bibinfo{volume}{47} (\bibinfo{year}{2024}) \bibinfo{pages}{102328}. \DOIprefix\doi{https://doi.org/10.1016/j.tsep.2023.102328}.
\bibitem[{Incropera et~al.(2007)Incropera, Dewitt, Bergman, and Lavine}]{incroperaFundamentalsHeatMass2007}
\bibinfo{author}{F.~P. Incropera}, \bibinfo{author}{D.~P. Dewitt}, \bibinfo{author}{T.~L. Bergman}, \bibinfo{author}{A.~S. Lavine}, \bibinfo{title}{Fundamentals of Heat and Mass Transfer}, \bibinfo{edition}{6} ed., \bibinfo{publisher}{{John Wiley \& Sons}}, \bibinfo{address}{Hoboken, USA}, \bibinfo{year}{2007}.
\bibitem[{Collier and Thome(1994)}]{Collier-1994-ConvBoilCond}
\bibinfo{author}{J.~Collier}, \bibinfo{author}{J.~Thome}, \bibinfo{title}{Convective boiling and condensation}, \bibinfo{edition}{3rd} ed., \bibinfo{publisher}{Oxford University Press}, \bibinfo{address}{New York}, \bibinfo{year}{1994}.
\bibitem[{Persad and Ward(2016)}]{persad-2016_Chem.Rev.}
\bibinfo{author}{A.~H. Persad}, \bibinfo{author}{C.~A. Ward},
\newblock \bibinfo{title}{Expressions for the evaporation and condensation coefficients in the {{Hertz}}-{{Knudsen}} relation},
\newblock \bibinfo{journal}{Chem. Rev.} \bibinfo{volume}{116} (\bibinfo{year}{2016}) \bibinfo{pages}{7727--7767}. \DOIprefix\doi{https://doi.org/10.1021/acs.chemrev.5b00511}.
\bibitem[{Stephan et~al.(1987)Stephan, Krauss, and Laesecke}]{stephan1987-Physical-chemical-ref}
\bibinfo{author}{K.~Stephan}, \bibinfo{author}{R.~Krauss}, \bibinfo{author}{A.~Laesecke},
\newblock \bibinfo{title}{Viscosity and thermal conductivity of nitrogen for a wide range of fluid states},
\newblock \bibinfo{journal}{Journal of physical and chemical reference data} \bibinfo{volume}{16} (\bibinfo{year}{1987}) \bibinfo{pages}{993--1023}. \DOIprefix\doi{https://doi.org/10.1063/1.555798}.
\bibitem[{Scott(1976)}]{scott1976-Physics_report}
\bibinfo{author}{T.~A. Scott},
\newblock \bibinfo{title}{Solid and liquid nitrogen},
\newblock \bibinfo{journal}{Physics Reports} \bibinfo{volume}{27} (\bibinfo{year}{1976}) \bibinfo{pages}{89--157}.
\bibitem[{Cook and Davey(1976)}]{cook-1976_Cryogenicsa}
\bibinfo{author}{T.~Cook}, \bibinfo{author}{G.~Davey},
\newblock \bibinfo{title}{The density and thermal conductivity of solid nitrogen and carbon dioxide},
\newblock \bibinfo{journal}{Cryogenics} \bibinfo{volume}{16} (\bibinfo{year}{1976}) \bibinfo{pages}{363--369}. \DOIprefix\doi{https://doi.org/10.1016/0011-2275(76)90217-4}.
\bibitem[{Flynn(2004)}]{flynn2004cryogenic}
\bibinfo{author}{T.~Flynn}, \bibinfo{title}{Cryogenic engineering}, \bibinfo{publisher}{CRC Press}, \bibinfo{year}{2004}.
\bibitem[{Lemmon et~al.(2007)Lemmon, Huber, and McLinden}]{lemmon2007nist}
\bibinfo{author}{E.~Lemmon}, \bibinfo{author}{M.~L. Huber}, \bibinfo{author}{M.~O. McLinden}, \bibinfo{title}{{NIST} standard reference database 23: reference fluid thermodynamic and transport properties-{{REFPROP}}, version 8.0}, \bibinfo{year}{2007}.
\bibitem[{Arp et~al.(2005)Arp, McCarty, and Jeffrey}]{arp2005-INC}
\bibinfo{author}{V.~Arp}, \bibinfo{author}{R.~McCarty}, \bibinfo{author}{F.~Jeffrey}, \bibinfo{title}{{HEPAK}}, \bibinfo{year}{2005}.
\bibitem[{Van~Sciver(2012)}]{van_sciver_helium_2012}
\bibinfo{author}{S.~W. Van~Sciver}, \bibinfo{title}{Helium Cryogenics}, International cryogenics monograph series, \bibinfo{edition}{2} ed., \bibinfo{publisher}{Springer}, \bibinfo{address}{New York, USA}, \bibinfo{year}{2012}.
\bibitem[{Bao and Guo(2021)}]{Bao-2021-PRB}
\bibinfo{author}{S.~R. Bao}, \bibinfo{author}{W.~Guo},
\newblock \bibinfo{title}{Transient heat transfer of superfluid $^{4}\mathrm{He}$ in nonhomogeneous geometries: Second sound, rarefaction, and thermal layer},
\newblock \bibinfo{journal}{Phys. Rev. B} \bibinfo{volume}{103} (\bibinfo{year}{2021}) \bibinfo{pages}{134510}. \DOIprefix\doi{https://doi.org/10.1103/PhysRevB.103.134510}.
\bibitem[{Sanavandi et~al.(2022)Sanavandi, Hulse, Bao, Tang, and Guo}]{Sanavandi-2022-PRB}
\bibinfo{author}{H.~Sanavandi}, \bibinfo{author}{M.~Hulse}, \bibinfo{author}{S.~R. Bao}, \bibinfo{author}{Y.~Tang}, \bibinfo{author}{W.~Guo},
\newblock \bibinfo{title}{Boiling and cavitation caused by transient heat transfer in superfluid helium-4},
\newblock \bibinfo{journal}{Phys. Rev. B} \bibinfo{volume}{106} (\bibinfo{year}{2022}) \bibinfo{pages}{054501}. \DOIprefix\doi{https://doi.org/10.1103/PhysRevB.106.054501}.
\bibitem[{Smith(1969)}]{smith-1969_Cryogenicsa}
\bibinfo{author}{R.~Smith},
\newblock \bibinfo{title}{Review of heat transfer to helium {{I}}},
\newblock \bibinfo{journal}{Cryogenics} \bibinfo{volume}{9} (\bibinfo{year}{1969}) \bibinfo{pages}{11--19}. \DOIprefix\doi{https://doi.org/10.1016/0011-2275(69)90251-3}.
\bibitem[{Lantz(2007)}]{lantz_heat_2007}
\bibinfo{author}{J.~Lantz}, \bibinfo{title}{Heat transfer correlations between a heated surface and liquid superﬂuid helium}, \bibinfo{type}{Thesis}, Linköpings universitet, \bibinfo{address}{Linköping, Sweden}, \bibinfo{year}{2007}.
\bibitem[{Zuber et~al.(1961)Zuber, Tribus, and Westwater}]{zuber_hydrodynamic_1961}
\bibinfo{author}{N.~Zuber}, \bibinfo{author}{M.~Tribus}, \bibinfo{author}{J.~Westwater},
\newblock \bibinfo{title}{The hydrodynamic crisis in pool boiling of saturated and subcooled liquids},
\newblock in: \bibinfo{booktitle}{Proceeding of 1961-62 International Heat Transfer Conference}, volume~\bibinfo{volume}{2} of \textit{\bibinfo{series}{International development in heat transfer}}, \bibinfo{publisher}{ASME}, \bibinfo{address}{Boulder, USA}, \bibinfo{year}{1961}, pp. \bibinfo{pages}{230--236}.
\bibitem[{Breen and Westwater(1962)}]{breen_effect_1962}
\bibinfo{author}{B.~P. Breen}, \bibinfo{author}{J.~W. Westwater},
\newblock \bibinfo{title}{Effect of diameter of horizontal tubes on film boiling heat transfer},
\newblock \bibinfo{journal}{Chemical Engineering Progress} \bibinfo{volume}{58} (\bibinfo{year}{1962}) \bibinfo{pages}{67--72}.
\bibitem[{Vinen(1957)}]{Vinen-1957-PRS-III}
\bibinfo{author}{W.~F. Vinen},
\newblock \bibinfo{title}{Mutual friction in a heat current in liquid helium {{II III}}. {{Theory}} of the mutual friction},
\newblock \bibinfo{journal}{Proc. R. Soc. Lond. A} \bibinfo{volume}{242} (\bibinfo{year}{1957}) \bibinfo{pages}{493--515}. \DOIprefix\doi{https://doi.org/10.1098/rspa.1957.0191}.
\bibitem[{Schwarz(1978)}]{Schwarz-1978-PRB}
\bibinfo{author}{K.~W. Schwarz},
\newblock \bibinfo{title}{Turbulence in superfluid helium: Steady homogeneous counterflow},
\newblock \bibinfo{journal}{Phys. Rev. B} \bibinfo{volume}{18} (\bibinfo{year}{1978}) \bibinfo{pages}{245--262}. \DOIprefix\doi{https://doi.org/10.1103/PhysRevB.18.245}.
\bibitem[{Martin and Tough(1983)}]{Martin-1983-PRB}
\bibinfo{author}{K.~P. Martin}, \bibinfo{author}{J.~T. Tough},
\newblock \bibinfo{title}{Evolution of superfluid turbulence in thermal counterflow},
\newblock \bibinfo{journal}{Phys. Rev. B} \bibinfo{volume}{27} (\bibinfo{year}{1983}) \bibinfo{pages}{2788--2799}. \DOIprefix\doi{https://doi.org/10.1103/PhysRevB.27.2788}.
\bibitem[{Bewley et~al.(2008)Bewley, Paoletti, Sreenivasan, and Lathrop}]{Bewley-2008-PNAS}
\bibinfo{author}{G.~P. Bewley}, \bibinfo{author}{M.~S. Paoletti}, \bibinfo{author}{K.~R. Sreenivasan}, \bibinfo{author}{D.~P. Lathrop},
\newblock \bibinfo{title}{Characterization of reconnecting vortices in superfluid helium},
\newblock \bibinfo{journal}{Proc. Natl. Acad. Sci. U.S.A.} \bibinfo{volume}{105} (\bibinfo{year}{2008}) \bibinfo{pages}{13707--13710}. \DOIprefix\doi{https://doi.org/10.1073/pnas.0806002105}.
\bibitem[{Chagovets and Van~Sciver(2011)}]{Chagovets-2011-PF}
\bibinfo{author}{T.~V. Chagovets}, \bibinfo{author}{S.~W. Van~Sciver},
\newblock \bibinfo{title}{A study of thermal counterflow using particle tracking velocimetry},
\newblock \bibinfo{journal}{Phys. Fluids} \bibinfo{volume}{23} (\bibinfo{year}{2011}) \bibinfo{pages}{107102}. \DOIprefix\doi{https://doi.org/10.1063/1.3657084}.
\bibitem[{Mineda et~al.(2013)Mineda, Tsubota, Sergeev, Barenghi, and Vinen}]{Mineda-2013-PRB}
\bibinfo{author}{Y.~Mineda}, \bibinfo{author}{M.~Tsubota}, \bibinfo{author}{Y.~A. Sergeev}, \bibinfo{author}{C.~F. Barenghi}, \bibinfo{author}{W.~F. Vinen},
\newblock \bibinfo{title}{Velocity distributions of tracer particles in thermal counterflow in superfluid ${}^{4}he$},
\newblock \bibinfo{journal}{Phys. Rev. B} \bibinfo{volume}{87} (\bibinfo{year}{2013}) \bibinfo{pages}{174508}. \DOIprefix\doi{https://doi.org/10.1103/PhysRevB.87.174508}.
\bibitem[{\ifmmode \check{S}\else \v{S}\fi{}van\ifmmode~\check{c}\else \v{c}\fi{}ara et~al.(2018)\ifmmode \check{S}\else \v{S}\fi{}van\ifmmode~\check{c}\else \v{c}\fi{}ara, Hrubcov\'a, Rotter, and La~Mantia}]{Svancara-2018-PRF}
\bibinfo{author}{P.~\ifmmode \check{S}\else \v{S}\fi{}van\ifmmode~\check{c}\else \v{c}\fi{}ara}, \bibinfo{author}{P.~Hrubcov\'a}, \bibinfo{author}{M.~Rotter}, \bibinfo{author}{M.~La~Mantia},
\newblock \bibinfo{title}{Visualization study of thermal counterflow of superfluid helium in the proximity of the heat source by using solid deuterium hydride particles},
\newblock \bibinfo{journal}{Phys. Rev. Fluids} \bibinfo{volume}{3} (\bibinfo{year}{2018}) \bibinfo{pages}{114701}. \DOIprefix\doi{https://doi.org/10.1103/PhysRevFluids.3.114701}.
\bibitem[{Marakov et~al.(2015)Marakov, Gao, Guo, Van~Sciver, Ihas, McKinsey, and Vinen}]{Marakov-2015-PRB}
\bibinfo{author}{A.~Marakov}, \bibinfo{author}{J.~Gao}, \bibinfo{author}{W.~Guo}, \bibinfo{author}{S.~W. Van~Sciver}, \bibinfo{author}{G.~G. Ihas}, \bibinfo{author}{D.~N. McKinsey}, \bibinfo{author}{W.~F. Vinen},
\newblock \bibinfo{title}{Visualization of the normal-fluid turbulence in counterflowing superfluid $^4${{He}}},
\newblock \bibinfo{journal}{Phys. Rev. B} \bibinfo{volume}{91} (\bibinfo{year}{2015}) \bibinfo{pages}{094503}. \DOIprefix\doi{https://doi.org/10.1103/PhysRevB.91.094503}.
\bibitem[{Gao et~al.(2016)Gao, Guo, L'vov, Pomyalov, Skrbek, Varga, and Vinen}]{Gao-2016-JETP}
\bibinfo{author}{J.~Gao}, \bibinfo{author}{W.~Guo}, \bibinfo{author}{V.~S. L'vov}, \bibinfo{author}{A.~Pomyalov}, \bibinfo{author}{L.~Skrbek}, \bibinfo{author}{E.~Varga}, \bibinfo{author}{W.~F. Vinen},
\newblock \bibinfo{title}{Decay of counterflow turbulence in superfluid ${^4}${He}},
\newblock \bibinfo{journal}{JETP Lett.} \bibinfo{volume}{103} (\bibinfo{year}{2016}) \bibinfo{pages}{648--652}. \DOIprefix\doi{https://doi.org/10.1134/S0021364016100064}.
\bibitem[{Bao and Guo(2019)}]{Bao-2019_PRApplied}
\bibinfo{author}{S.~Bao}, \bibinfo{author}{W.~Guo},
\newblock \bibinfo{title}{Quench-spot detection for superconducting accelerator cavities via flow visualization in superfluid helium-4},
\newblock \bibinfo{journal}{Phys. Rev. Applied} \bibinfo{volume}{11} (\bibinfo{year}{2019}) \bibinfo{pages}{044003}. \DOIprefix\doi{https://doi.org/10.1103/PhysRevApplied.11.044003}.
\bibitem[{Murakami et~al.(2015)Murakami, Takada, and Nozawa}]{Murakami-2015-IOP}
\bibinfo{author}{M.~Murakami}, \bibinfo{author}{S.~Takada}, \bibinfo{author}{M.~Nozawa},
\newblock \bibinfo{title}{Study of {He II} boiling flow field around a heater},
\newblock \bibinfo{journal}{{IOP} Conf. Ser.: Mater. Sci. Eng.} \bibinfo{volume}{101} (\bibinfo{year}{2015}) \bibinfo{pages}{012165}. \DOIprefix\doi{https://doi.org/10.1088/1757-899x/101/1/012165}.
\bibitem[{Claudet and Seyfert(1981)}]{Claudet-1981-ACE}
\bibinfo{author}{G.~Claudet}, \bibinfo{author}{P.~Seyfert},
\newblock \bibinfo{title}{Bath cooling with subcooled superfluid helium},
\newblock \bibinfo{journal}{{Adv. Cryog. Eng.}} \bibinfo{volume}{27} (\bibinfo{year}{1981}) \bibinfo{pages}{441}.
\bibitem[{Kashani and {Van Sciver}(1985)}]{Kashani-1985-Cryo}
\bibinfo{author}{A.~Kashani}, \bibinfo{author}{S.~W. {Van Sciver}},
\newblock \bibinfo{title}{High heat flux {Kapitza} conductance of technical copper with several different surface preparations},
\newblock \bibinfo{journal}{Cryogenics} \bibinfo{volume}{25} (\bibinfo{year}{1985}) \bibinfo{pages}{238--242}. \DOIprefix\doi{https://doi.org/10.1016/0011-2275(85)90202-4}.
\bibitem[{Van~Sciver and Lee(1980)}]{VanSciver-1980-ACE}
\bibinfo{author}{S.~W. Van~Sciver}, \bibinfo{author}{R.~Lee},
\newblock \bibinfo{title}{Heat transfer to helium-{II} in cylindrical geometries},
\newblock in: \bibinfo{booktitle}{Adv. Cryog. Eng.}, volume~\bibinfo{volume}{35}, \bibinfo{publisher}{{Springer US}}, \bibinfo{address}{Boston, MA.}, \bibinfo{year}{1980}, pp. \bibinfo{pages}{363--371}. \DOIprefix\doi{https://doi.org/10.1007/978-1-4613-9856-1\_43}.
\bibitem[{Goodling and Irey(1969)}]{Goodling-1969-ACE}
\bibinfo{author}{J.~S. Goodling}, \bibinfo{author}{R.~K. Irey},
\newblock \bibinfo{title}{Non-boiling and film boiling heat transfer to a saturated bath of liquid helium},
\newblock in: \bibinfo{booktitle}{Adv. Cryog. Eng.}, volume~\bibinfo{volume}{14}, \bibinfo{publisher}{{Springer US}}, \bibinfo{address}{{Boston, MA}}, \bibinfo{year}{1969}, pp. \bibinfo{pages}{159--169}.
\bibitem[{Bretts and Leonard(1975)}]{Bretts-1975-ACE}
\bibinfo{author}{K.~Bretts}, \bibinfo{author}{A.~Leonard},
\newblock \bibinfo{title}{Free convection film boiling from a flat, horizontal surface in saturated {{He II}}},
\newblock in: \bibinfo{booktitle}{Adv. in {{Cryog. Eng.}}}, volume~\bibinfo{volume}{21}, \bibinfo{publisher}{{Springer US}}, \bibinfo{address}{Boston, MA.}, \bibinfo{year}{1975}, pp. \bibinfo{pages}{282--292}. \DOIprefix\doi{https://doi.org/10.1007/978-1-4757-0208-8\_37}.
\bibitem[{Danaila et~al.(2007)Danaila, Joly, Kaber, and Postel}]{danaila_introduction_2007}
\bibinfo{author}{I.~Danaila}, \bibinfo{author}{P.~Joly}, \bibinfo{author}{S.~M. Kaber}, \bibinfo{author}{M.~Postel}, \bibinfo{title}{An introduction to scientific computing: Twelve computational projects solved with {MATLAB}}, \bibinfo{publisher}{Springer-Verlag}, \bibinfo{address}{New York, USA}, \bibinfo{year}{2007}.
\bibitem[{Sod(1978)}]{sod_survey_1978}
\bibinfo{author}{G.~A. Sod},
\newblock \bibinfo{title}{A survey of several finite difference methods for systems of nonlinear hyperbolic conservation laws},
\newblock \bibinfo{journal}{Journal of Computational Physics} \bibinfo{volume}{27} (\bibinfo{year}{1978}) \bibinfo{pages}{1--31}. \DOIprefix\doi{https://doi.org/10.1016/0021-9991(78)90023-2}.
\bibitem[{Darve et~al.(2014)Darve, Bosland, Devanz, Olivier, Renard, and Thermeau}]{darve2014ess}
\bibinfo{author}{C.~Darve}, \bibinfo{author}{P.~Bosland}, \bibinfo{author}{G.~Devanz}, \bibinfo{author}{G.~Olivier}, \bibinfo{author}{B.~Renard}, \bibinfo{author}{J.-P. Thermeau},
\newblock \bibinfo{title}{The {ESS} elliptical cavity cryomodules},
\newblock in: \bibinfo{booktitle}{AIP Conference Proceedings}, volume \bibinfo{volume}{1573}, \bibinfo{organization}{American Institute of Physics}, \bibinfo{year}{2014}, pp. \bibinfo{pages}{639--646}. \DOIprefix\doi{https://doi.org/10.1063/1.4860762}.
\bibitem[{Dylla(1994)}]{dylla1994development}
\bibinfo{author}{H.~Dylla},
\newblock \bibinfo{title}{Development of ultrahigh vacuum technology for particle accelerators and magnetic fusion devices},
\newblock \bibinfo{journal}{Journal of Vacuum Science \& Technology A: Vacuum, Surfaces, and Films} \bibinfo{volume}{12} (\bibinfo{year}{1994}) \bibinfo{pages}{962--978}. \DOIprefix\doi{https://doi.org/10.1116/1.579074}.
\bibitem[{Padamsee(2009)}]{padamseerf}
\bibinfo{author}{H.~Padamsee}, \bibinfo{title}{{RF} superconductivity: Science, technology, and applications (2009 wiley}, \bibinfo{year}{2009}.
\bibitem[{Padamsee et~al.(2008)Padamsee, Knobloch, and Hays}]{padamsee2008rf}
\bibinfo{author}{H.~Padamsee}, \bibinfo{author}{J.~Knobloch}, \bibinfo{author}{T.~Hays}, \bibinfo{title}{{RF} superconductivity for accelerators}, \bibinfo{publisher}{John Wiley \& Sons}, \bibinfo{year}{2008}.
\bibitem[{Garceau(2022)}]{garceau2022heat}
\bibinfo{author}{N.~Garceau}, \bibinfo{title}{Heat and Mass Transfer during a Sudden Loss of Vacuum in a Liquid Helium Cooled Tube}, Ph.D. thesis, The Florida State University, \bibinfo{year}{2022}.
\bibitem[{Launder(1972)}]{launder1972lectures}
\bibinfo{author}{B.~Launder}, \bibinfo{title}{Lectures in mathematical models of turbulence}, \bibinfo{year}{1972}.

\end{thebibliography}

\section*{Nomenclature}


\begin{tabular}[ht]{p{1cm}p{4cm}p{2.1cm}}
	\hline
	Variable 			& Description										   & Units \\
	\hline
	$B_w$       		& Coefficient in the film boiling correlation of He I  & W/(cm$^2\cdot$ K$^{5/4}$)    \\
 $c_0$& Speed of sound& m/s\\
	$C$       	    	& Specific heat                                        & J/(kg$\cdot $K)    \\
	$D_1$           	& Inner diameter of the tube                           & m                  \\
	$D_2$           	& Outer diameter of the tube                           & m                  \\
 $E$& Specific internal
energy& J/kg               \\
	$\hat{h}$           & Specific enthalpy                                    & J/kg               \\
	$H$           		& Immersion depth                                      & m	                \\
	$k$             	& Thermal conductivity                                 & W/(m$\cdot $K)     \\
	$\dot m_c$   & Mass deposition rate                                 & kg/(m$^2\cdot $s)  \\
	$\dot m$   			& Mass flow rate                                 	   & kg/s			    \\
	$M_g$             	& Gas molar mass                                       & kg/mol             \\
	$Nu$            	& Nusselt number                                       &                    \\
	$P$             	& Nitrogen gas pressure                                & Pa                 \\
 $Pr$& Prandtl number& \\
	$q$           		& Heat flux											   & W/m$^2$            \\
	$q^*$           	& Peak heat flux                                       & W/m$^2$            \\
	$q_{con}$      & Conductive heat flux along the wall                  & W/m$^2$            \\
	$q_{He}$       & Heat flux to the liquid helium bath                  & W/m$^2$            \\
	$R$             	& Ideal gas constant                                   & J/(mol$\cdot $K)   \\
 $Re$& Reynolds number& \\
	$t$             	& Time                                                 & s                  \\
	$t_r$             	& Arrival time of gas front                            & s                  \\
	$T$             	& Temperature                                          & K                  \\
	$T_b$        & Helium bath temperature                              & K                  \\
	$T_c$        & Center temperature of the SN$_2$ layer               & K                  \\
 $u$& Radial mean velocity& m/s                \\
 $u_T$& Gas molecular thermal velocity& m/s                \\
	$v$             	& Gas velocity                                         & m/s                \\
	$v_0$&  Escape speed& m/s                \\
	$x$             	& Axial coordinate                                     & m                  \\
	$x_F$& Freeze range                   	                   & m                  \\
	
						&                                                      &                    \\
	$Greeks$        	&                                                      &                    \\
 $\beta$& Ratio of velocities $u/u_T$& \\
 $\gamma$& Ratio of the specific heats& \\
	$\Gamma$        	& Schrage parameter                                    &                    \\
	$\delta$        	& Thickness of the SN$_2$ layer                        & m                  \\
 $\delta_g$& Height of the grid cell&m\\
	$\dot{\delta}$      & Change rate of $\delta$                        	   & m/s                \\
	$\Delta T$        	& Temperature difference                        	   & K                  \\
	$\varepsilon$   	& Specific internal energy                             & J/kg               \\
 $\mu$& Viscosity& Pa$\cdot $s\\
	$\rho$          	& Density                                              & kg/m$^3$           \\
\end{tabular}

\begin{tabular}[ht]{p{1cm}p{4cm}p{2.1cm}}
\hline
 $\sigma_c$& Condensation coefficient
&\\
 $\sigma_e$& Evaporation coefficient                              &\\
	
 $\tau$& Stress tensor& kg/(m$\cdot$s$^2$)\\
	$\Psi$          	& Parameter in the integral correlation of $q^*$ for He II                & kg/(m$^2\cdot $s)  \\
						&                                                      &                    \\
	$Subscripts$    	&                                                      &                    \\
 $eff$& Effective& \\
	$g$          & Bulk gas condition                                   &                    \\
	$s$          & Surface of SN$_2$ layer           				   &                    \\
	$w$          & Copper tube wall                                     &                    \\
	$SN$         & Solid nitrogen                                       &                    \\
	\hline
\end{tabular}

\end{document}